\documentstyle[12pt]{article}
\input epsf
\begin{document}
\newcommand{\gtrsim}{\,\raisebox{-0.13cm}{$\stackrel{\textstyle>}
{\textstyle\sim}$}\,}
\newcommand{\lessim}{\,\raisebox{-0.13cm}{$\stackrel{\textstyle<}
{\textstyle\sim}$}\,}

\rightline{RU-94-04}
\rightline{January 1994}
\baselineskip=18pt
\vskip 0.5in
\begin{center}
{\bf \LARGE Radiative Decay of Vector Quarkonium:\\
Constraints on Glueballs\\ and Light Gluinos } \\
\vspace*{0.5in}
{\large Mesut Bahad{\i}r \c{C}ak{\i}r}\footnote{Present address: 150
Fitzrandolph Rd,
Princeton, NJ 08540} and
{\large Glennys R. Farrar}\footnote{Research supported
in part by NSF-PHY-91-21039} \\
\vspace{.15in}
{\it Department of Physics and Astronomy \\ Rutgers University,
Piscataway, NJ 08855, USA}

\end{center}
\vskip  0.2in
{\bf Abstract:}
Given a resonance of known mass, width, and $J^{PC}$, we can determine
its gluonic branching fraction, $b_{R\rightarrow g g}$, from data on
its production in radiative vector quarkonium decay, $V \rightarrow
\gamma R$.  For most resonances $b_{R\rightarrow g g}$ is found to be
$O(10\%)$, consistent with being $q \bar{q}$ states, but we find that
both pseudoscalars observed in the 1440 MeV region have
$b_{R\rightarrow g g} \sim \frac{1}{2} - 1$, and $b(f_{0^{++}}
\rightarrow gg)\sim \frac{1}{2}$.  As data improves, $b_{R\rightarrow
g g}$ should be a useful discriminator between $q \bar{q}$ and gluonic
states and may permit quantitative determination of the extent to
which a particular resonance is a mixture of glueball and $q\bar{q}$.
We also examine the regime of validity of pQCD for predicting the rate
of $V\rightarrow \gamma \eta_{\tilde{g}}$, the ``extra'' pseudoscalar
bound state which would exist if there were light gluinos.  From the
CUSB limit on peaks in $\Upsilon \rightarrow \gamma X$, the mass range
3 GeV $ \lessim m(\eta_{\tilde{g}}) \lessim $ 7 GeV can be excluded.
An experiment must be significantly more sensitive
to exclude an $\eta_{\tilde{g}}$ lighter than this.

\thispagestyle{empty}
\newpage
\section{Introduction}
\label{sec:intro}
\begin{figure}
\epsfxsize=\hsize
\epsffile{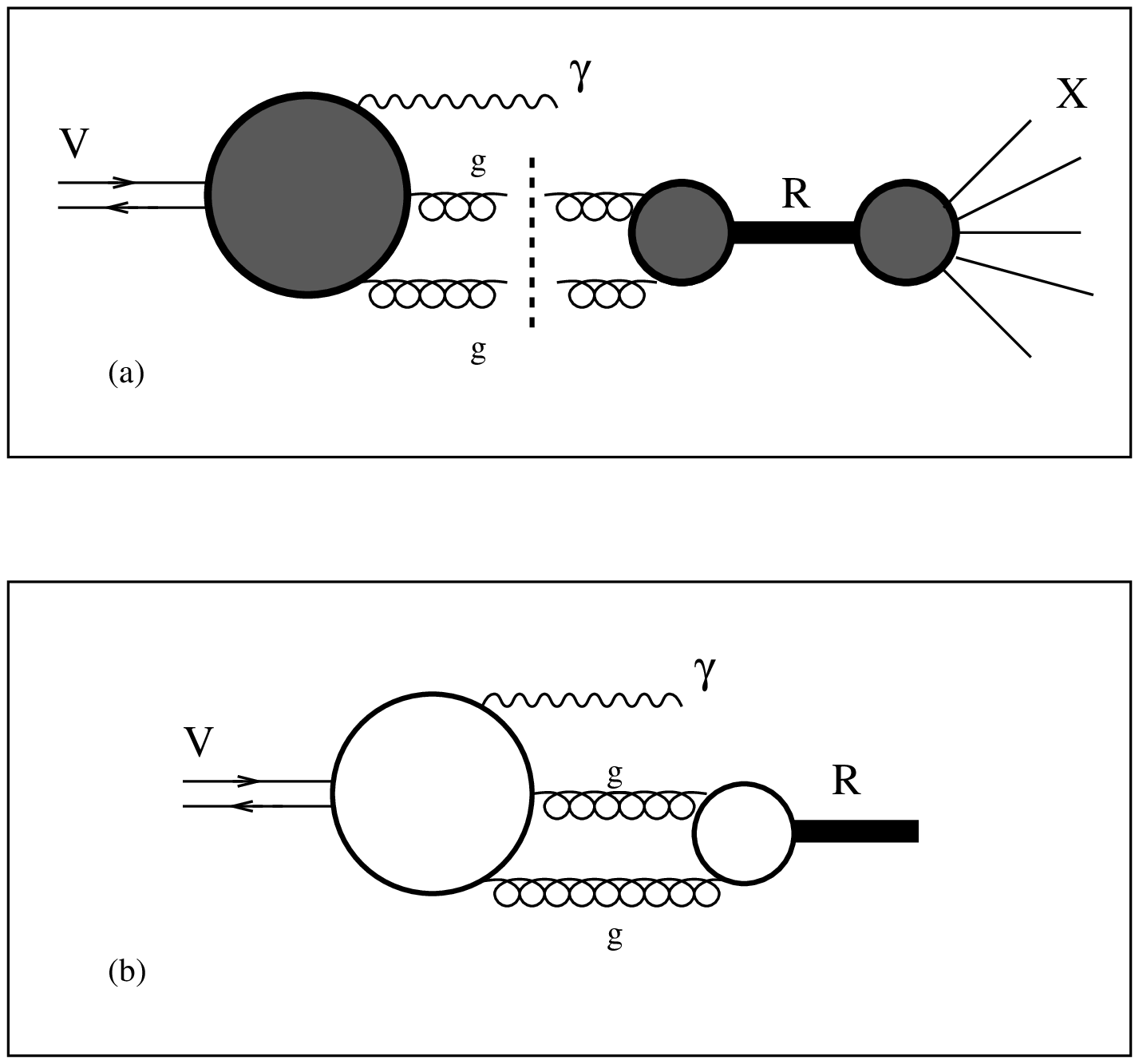}
\caption{The dominant contribution to $ V \rightarrow \gamma R$.}
\label{VR.ps}
\end{figure}

\hspace*{2em}
Radiative quarkonium decay has been experimentally studied both
exclusively and inclusively.  It is a particularly auspicious reaction
for producing glueball resonances, as is evident from
Fig. \ref{VR.ps}.  In exclusive channels such as $J/\Psi \rightarrow
\gamma K\bar{K}\pi$, many resonances in the $K\bar{K}\pi$ invariant
mass have been observed.  In the inclusive process
$\Upsilon\rightarrow \gamma X$, the absence of peaks in the photon
energy spectrum provides an upper bound on the production of
resonances\cite{tutsmunich,cusb}:\footnote{The actual limit is
slightly dependent on the mass of the resonance, but varies less than
a factor of 2 as the mass varies between 1.5 GeV and 8 GeV.  The exact
value of the limit doesn't matter much when the resonance mass is
large, as will be seen below, so we take the value in the range which
is most relevant to analyzing data, $\sim 2$ GeV.}
\begin{equation}
\frac {\Gamma(\Upsilon\rightarrow\gamma R)}{\Gamma(\Upsilon\rightarrow all)}
   \lessim 8\times 10^{-5}.
\label{cusb_bound}
\end{equation}
Qualitatively, a resonance which is prominent in $V\rightarrow\gamma
R$ but which is not prominent in hadronic scattering, can be a good
candidate for a glueball.  More quantitatively,
``stickiness"\cite{chanowitz}, which is proportional to
$\Gamma(V\rightarrow\gamma R)$ divided by the 2-photon partial width,
$\Gamma(R \rightarrow \gamma \gamma)$, should be significantly larger for
glueballs than for $q \bar{q}$ mesons.  In the following we propose
another quantitative measure of the gluonic content of a resonance:
its branching ratio to gluons, $b_{R\rightarrow gg}$.  We describe how
to extract it from experimental information on the width of the
resonance and the branching ratio for its production in radiative
quarkonium decay, $b(V\rightarrow\gamma R)\equiv
\Gamma(\Upsilon\rightarrow\gamma R)/\Gamma(\Upsilon\rightarrow all)$.
We work in a naive parton approximation, but expect that the method
can be shown to be more general.  We will see that it seems to be a
promising discriminator between glueballs and $q\bar{q}$ states, with
glueball candidates having $b_{R\rightarrow gg} \sim \frac{1}{2} - 1 $
and $q \bar{q}$ states having $b_{R\rightarrow gg} \sim \alpha_s^2
\sim 0.1 $.

If the resonance $R$ is a massive $Q \bar{Q}$ state such as the
$\eta_c$, the branching ratio $b(V \rightarrow \gamma R)$ can be
reliably calculated in pQCD with a non-relativistic potential for both
$V$ and $R$\cite{guberina_kuhn}.  However in the interesting cases of
$R$ being a glueball or a bound state of light gluinos\footnote{While
beamdump and collider experiments have ruled out light gluinos which
decay in the apparatus, longer lived gluinos are not ruled out, except
within certain ranges of mass and lifetime\cite{f:51,f:82}.
Long-lived gluinos would form hadrons, with the ground state
$\tilde{g}\tilde{g}$ being a $0^{-+}$.  Demonstrating that the hadron
spectrum cannot accomodate an additional flavor singlet pseudoscalar
would thus exclude long lived
gluinos\cite{keung_khare,kuhn_ono,goldman_haber}.}, no reliable means
exists for making an absolute prediction for $b(V \rightarrow \gamma
R)$.  Therefore up to now we have been unable to make quantitative use of the
exclusive data and the upper limit (\ref{cusb_bound}) to address the
question of whether a given resonance is a glueball, $q \bar{q}$ or
possibly $\tilde{g}\tilde{g}$ state.  This paper proposes a means to
rectify this situation.

Let us begin by reviewing the case that the $R$ is a bound state of massive
quarks and $V$ is the $\Upsilon$.  Since the mass of the $b$ quark is
large, pQCD is believed to be a reliable means of computing the decay
$\Upsilon\rightarrow\gamma gg$.  If the produced resonance, $R$, also
contains heavy quarks,then pQCD can be reliably used to compute the
branching ratio $b(\Upsilon\rightarrow\gamma R )$ in terms of $|R(0)|^2$,
the resonance wave function at short distance,\footnote{For other than
pseudoscalar resonances, the derivative of the short-distance wave function
enters.} $\alpha_s$ and $m_R$.  In practice, one takes a non-relativistic
model for the $Q\bar{Q}$ potential and fixes its parameters to give a
correct prediction for $\Gamma(\eta_c \rightarrow e^+ e^-)$, which also
depends on $|R(0)|^2 $.  Having fixed the parameters of the potential,
$|R(0)|^2$ for the other $Q\bar{Q}$ resonances is determined, assuming that
they are described by the same non-relativistic potential model.  Following
this procedure, the branching fractions $b(V\rightarrow\gamma R)$ have been
predicted for the known $Q\bar{Q}$ mesons\cite{guberina_kuhn,kkks} and are
found to be small enough that they would not have been seen in the CUSB
experiment.  Kuhn\cite{kuhn:lightquarks} showed how to obtain
$b(V\rightarrow\gamma R)$ when $R$ is a light quark meson, in terms of its
decay constant, e.g., $f_{\pi}$ for $R$ a pion.

If there were a gluino with mass $m_{\tilde{g}}$ large enough for the
non-relativistic potential description to be valid, we would have an extra
$O^{-+}$ resonance with $m_R \approx 2 m_{\tilde{g}}$\footnote{Of course it
can mix with nearby glueball or $q\bar{q}$ resonances, but we keep this
discussion simple and we neglect mixing.  It can be introduced with no
conceptual difficulty.}.  In this case, pQCD can be used to calculate
$b(V\rightarrow\gamma\eta_{\tilde{g}})$\cite{keung_khare,kuhn_ono,goldman_haber}.
The result is larger than for the $\eta_c$ by a factor $\sim 21.5$ due to
the gluino being a color octet rather than triplet as we will detail
below.  For $m_{\eta_{\tilde{g}}} \gtrsim 3$ GeV, the pQCD calculation with
non-relativistic potential is probably a good enough approximation that the
data can be used to exclude a gluino.

Of course the wavefunctions of glueballs and of mesons made of light
quarks (or gluinos) cannot be treated with a non-relativistic
potential model.  No-one would try to calculate the width of a
glueball of mass $\sim 1.5-2$, or even 3, GeV on the basis of knowing
the $\eta_c$ width!  Nevertheless, we shall see that (eq. (\ref{final}
below) $b(\Upsilon\rightarrow\gamma R)$ for $R$ a glueball, or light
$q\bar{q}$ or gluino resonance can be estimated in terms of the
observed width of the resonance, $\Gamma_R$, and its branching ratio
to gluons, $b_{R\rightarrow gg}$.  Thus for a given resonance whose
width has been measured, by comparing this prediction with the CUSB
limit we find an upper limit on $b_{R\rightarrow gg}$ for that
resonance.  For resonances observed in $J/\Psi\rightarrow\gamma
K\bar{K}\pi$, we can extract the product $b_{R\rightarrow gg} \cdot
b(R\rightarrow K\bar{K}\pi)$.  When $b(R\rightarrow K\bar{K}\pi)$ is
known, $b_{R\rightarrow gg}$ can be determined.  Otherwise the
requirement $b(R\rightarrow K\bar{K}\pi)\lessim 1$ provides a lower limit
on $b_{R\rightarrow gg}$. In some cases these upper and lower bounds
are quite close to one another as we shall see below.  If
$b_{R\rightarrow gg}\sim 1$, it is a viable glueball candidate, while
for a $q\bar{q}$ resonance we would expect $b_{R\rightarrow gg} \sim
\alpha_s^2 \sim O(1/10)$.  Mixing between $q\bar{q}$ and glueball
resonances will give intermediate values of $b_{R \rightarrow gg}$.
Although the data is still inadequate to draw firm conclusions, we
will see possible examples of all three cases.

Using the formalism developed below, the CUSB data can be turned
into an upper limit on the gluonic width of of any resonance produced
in $\Upsilon\rightarrow \gamma R$.  This is the best way to quote a
limit on a possible $\eta_{\tilde{g}}$, since if it is not an
already-observed resonance, its width is unknown.  The question then
becomes whether the allowed width is consistent with theoretical
expectations for its width, for a given mass.

\section{Unitarity Calculation of the Absorptive Contribution}
\label{sec:unitarity}
\hspace*{2em}
Since we want a method of computing $b(V \rightarrow \gamma R)$ which
is not limited to $R$ being a massive $Q\bar{Q}$ state, let us see how
far we can go with unitarity and analyticity.  We begin by computing
the ``unitarity lower bound" on $\Gamma(V\rightarrow\gamma R)$ coming
from the two gluon intermediate state, as illustrated in Fig. 1. This
is an idea which cannot be precisely defined, since the gluon is not
an asymptotic state.  Our procedure can be made more rigorous by
introducing a scale on which one sees two gluons in the intermediate
state and does not resolve them further.  This scale will be related
to the ir cutoff and uv renormalization which must be introduced when
working beyond tree approximation.  However we begin with the most
naive approach.  We take the gluons to be massless.  As a consistency
check that this naive approach is reasonable, we verify below that
giving the gluons masses $\sim \Lambda_{QCD}$ makes no significant
difference to the conclusions.  We will return later to the question
of multigluon contributions and the resolution-size dependence.

In this approximation, the absorptive part of ${{\cal M}^R}_{V
\rightarrow \gamma X}$ can be fixed in terms of the width $\Gamma(R
\rightarrow gg)$, the inclusive radiative decay rate $\Gamma(V
\rightarrow \gamma g g)$, and the rate $\Gamma(R \rightarrow X)$, as
follows.  Since we can safely ignore interactions between the photon
and the final resonance $R$, unitarity tells us
\begin{equation}
{\cal M}_{V \rightarrow \gamma X} (P, k, \{p_i\}) = \sum_{n}{\cal
        M}_{V \rightarrow \gamma n} (P, k) \; {\cal M}_{n \rightarrow
        X} (s, \{p_i\})
\label{unitarity}
\end{equation}
where $n$ labels the intermediate state and $P,~k,~ \{p_i\}$ are the
4-momenta of $V, ~\gamma$, and the final state hadrons in $X$.  $s$ is
the (invariant mass)$^2$ of the state $X$.  Although $J^{PC}$ of the
resonance is fixed, in general more than one $L,S$ state of two gluons
can contribute for a given $J^{PC}$.

Now we want to rewrite (\ref{unitarity}) as an integral
over $s$ by inserting
\begin{eqnarray}
1~=~
{\displaystyle \int ds\
                \delta^{(4)}(p-\sum_{i}p_{i})\ d^{4}p\ \delta (s-p^{2})}
{}~=~
{\displaystyle  \int  \frac{ds}{2\pi} \frac{d^{3}{\bf p}}{(2\pi)^3 2E _p}
                        \ (2\pi)^4\ \delta^{(4)}(p-\sum_{i} p_{i})}
\label{216}
\end{eqnarray}
where $p$ is the 4-momentum of the state $X$ and $E_p \equiv
\sqrt{{\bf p}^2 + s}$.  Assuming that just a single 2-gluon state
dominates, or if more than one is important that that their
contributions add incoherently, we obtain
\begin{eqnarray}
\Gamma^R_{V \rightarrow \gamma X} ~ = ~ \int \frac{ds}{2\pi}
        \left\{ \frac{1}{2M_V} \int \frac{d^3{\bf k}}{(2 \pi)^3 2E_k}
        \frac{d^3{\bf p}}{(2\pi)^3 2E_p} \ (2 \pi)^4 \delta^4 (P-k-p)
        |{\cal M}_{V \rightarrow \gamma n} (P,k)|^2
        \right\}
\nonumber \\
 \times
\left\{\prod_i \left(\frac{d^3{\bf p_i}}{(2 \pi)^3E_{p_{i}}}\right)
        (2 \pi)^4 \delta ^4(p-\sum_ip_{i})
                |{\cal M}^R_{n \rightarrow X} (s,\{p_i\})|^2
                \right\}
\nonumber \\
{}~ = ~ \int \frac{ds}{2\pi} \frac{d\Gamma_{V \rightarrow \gamma n}}{ds}
\left\{\prod_i \left(\frac{d^3{\bf p_i}}{(2 \pi)^3E_{p_i}}\right)
        (2 \pi)^4 \delta ^4(p-\sum_ip_{i})
                |{\cal M}^R_{n \rightarrow X} (s,\{p_i\})|^2
                \right\},
\label{width}
\end{eqnarray}
\\
where
\begin{eqnarray}
\frac{d\Gamma_{V \rightarrow \gamma n}}{ds}  ~\equiv ~
\frac{1}{2M_V}\int \frac{d^3{\bf k}}{(2 \pi)^3 2E_k} \
        \frac{d^3{\bf p}}{(2\pi)^3 2E_p} \ (2 \pi)^4 \delta^4 (P-k-p)
        |{\cal M}_{V \rightarrow \gamma n} (P,k)|^2.
\label{diffl}
\end{eqnarray}
We define ${\cal P}^J_{S;L}$ to be the probability that the two gluons with
spins $s_1$ and $s_2$ in a total spin state $S$ and having orbital
angular momentum $L$ will combine to form the state with the $J^{PC}$
of the resonance $R$.  Then
\begin{equation}
\frac{d\Gamma_{V \rightarrow \gamma n}}{ds}
{}~=~\frac{d\Gamma_{V \rightarrow \gamma gg}}{ds} {\cal P}^J_{S;L}.
\end{equation}
$\frac{d\Gamma_{V\rightarrow\gamma gg}}{ds}$ can in principle be taken
directly from the measured inclusive radiative decay spectrum or, due
to the heavy mass of the quark in $V$, should be reliably given by
pQCD. If the latter route is taken, one would just project onto the
relevant $J^{PC}$ for the two gluons and automatically include the
correct ${\cal P}^J_{S;L}$\cite{kkks}.  Although within error bars the
data on $\frac{d\Gamma_{V\rightarrow\gamma X}}{ds}$ agrees with the
pQCD predictions, we adopt here the pQCD approach since the data on
$\frac{d\Gamma_{V\rightarrow\gamma X}}{ds}$ has large error bars for
the $s$ range of greatest interest, and projection onto the correct
$J^{PC}$ state of the 2-gluons is most reliably done using
pQCD.\footnote{See \cite{cakir} for a detailed treatment when data
rather than pQCD is used.}

Returning to (\ref{width}), we require the matrix element ${\cal
M}^R_{n\rightarrow X}$.  For $s$ near $m_R^2$ it is given by the
Breit-Wigner expression
\begin{equation}
{\cal{M}}^{R}_{n \rightarrow X} \equiv
{\displaystyle \frac {{\cal {M}}_{n \rightarrow R}
{\cal{M}}_{R \rightarrow X}}
{(s-m_{R}^{2})\ +\ im_{R}\Gamma_{R}}}\ ,
\label{BW}
\end{equation}
where $\Gamma_{R}$ is the total decay width of the resonance.  Now,
neglecting possible variation of the matrix elements with $s$ over the
width of the resonance, the expression in curly brackets in
(\ref{width}) is just:
\begin{equation}
{\displaystyle \left\{ \ \ \right\}} =
                {\displaystyle \left\{ \frac{1}{\Phi_n(s)}
        \left( \frac{4\ m_{R}^{2}\  \Gamma_{R \rightarrow n}
\Gamma_{R \rightarrow X}}
{(s-m_{R}^{2})^{2}\ +\ m_{R}^{2}\ \Gamma_{R}^{2}}\right) \right\}},
\label{eq:nRX}
\end{equation}
where $m_{R}$ is the mass of the resonance and $\Gamma_{R \rightarrow
n}$, $\Gamma_{R \rightarrow X}$ and $\Gamma_R$ are its partial and
total decay widths in obvious notation.  Equation (\ref{eq:nRX})
follows since\\
\begin{equation}
{\displaystyle \int \prod_{i}
        \left[\frac{d^3{\bf p_i}}{(2\pi)^{3}2E_{i}} \right]}
        (2\pi)^{4}\ \delta^{(4)}(p-{\displaystyle\sum_{i}} p_{i})
        \mid {\cal{M}}_{R \rightarrow X} \mid^{2}~ =~ 2\ m_{R}\ \Gamma_{R
\rightarrow X}
\end{equation}
\\
and, in the approximation that $\mid {\cal{M}}_{n \rightarrow R} \mid^{2} $
is approximately constant over the resonance,\\
\begin{equation}
\mid {\cal{M}}_{n \rightarrow R} \mid^{2} =
        {\displaystyle \frac{2\ m_{R}}{\Phi_{n}(s)}\ \Gamma_{R \rightarrow n}}.
\label{eq:MnR}
\end{equation}
\\
The phase space factor $\Phi_n(s)$ is $\frac{\lambda (s)}{8\pi}$, with\\
\begin{equation}
\lambda (s)~\equiv~
        \frac{1}{s}\sqrt{s^2+m_1^4+m_2^4-2m_1^2 m_2^2-2s~ m_1^2-2s~ m_2^2}
\end{equation}
\\
for a two particle intermediate state. For real gluons $m_1=m_2=0$ and
$\lambda (s)=1$, while for $m_1=m_2= 200~(500)$ MeV, $\lambda((1.5
{\rm GeV})^2)= 0.96~(0.75)$.  Thus taking gluons to be massless or to
have masses $O(\Lambda_{QCD})$ makes only a small difference in the
conclusions\footnote{Except when $R$ is a $1^{++}$, in which case the
leading absorptive amplitude vanishes for strictly massless gluons and
the more rigorous treatment introducing a scale would be essential to
obtaining a reliable result.}, so we simply set $\lambda(s) = 1$
hereafter.

Putting together eqns. (\ref{width}) and (\ref{eq:nRX}), we obtain the
absorptive contribution to the rate:
\begin{equation}
\Gamma^{R~(abs)}_{V \rightarrow \gamma X}~=~ 16 \pi \ {\cal P}^J_{S;L}
        b_{R \rightarrow gg} b_{R \rightarrow X}
\int \frac{ds}{\pi\lambda(s)} \frac{d\Gamma_{V\rightarrow\gamma gg}}{ds}
        \frac{ m^2_R\Gamma^2_R }{(s-m_R^2)^2+m_R^2\Gamma_R^2}\ .
\end{equation}
\\
In the narrow width limit\\
\begin{equation}
{\displaystyle \lim_{\Gamma \rightarrow 0}\
\left[\frac{m\ \Gamma/\pi}{(s-m^{2})^{2}\ +\ m^{2}\ \Gamma^{2}} \right]}
                ~=~ \delta(s-m^{2})\  .
\end{equation}
As a result, the absorptive contribution to the branching ratio for
$V\rightarrow \gamma R$ in the narrow resonance limit is:\\
\begin{equation}
b_{V\rightarrow \gamma R}^{(abs)} ~=~16 \pi~
        m_R\Gamma_R {\cal P}^J_{S;L} \left( \frac{1}{\Gamma_V}
        \frac{d\Gamma_{V\rightarrow \gamma gg}}{ds}\Big{|}_{s=m^2_R}\right)
        b_{R\rightarrow gg},
\label{B}
\end{equation}
where $\Gamma_{V}$ is the total width of the initial
vector meson and $b_{R\rightarrow gg}$ is the gluonic branching
fraction of $R$.\footnote{Nominally, it is the 2-gluon branching
fraction, but when the scale of resolution of the gluons is taken
large enough, as is implicit in our partonic discussion, this is just
the total gluonic width of $R$.}

Note that even when $m_R$ is $O(1$ GeV) our unitarity calculation with
a two-gluon intermediate state is a good approximation.  This is
because the quarks in $V$ are quite massive so that three hard gluons
are suppressed by $O(\alpha_s(m_b))$ in their contribution to $\Gamma(
V \rightarrow \gamma X)$ compared to that of two gluons.  Moreover in
the partonic spirit of this paper, by taking the resolution size of
the ``effective" gluons we are discussing to be large enough, we can
arrange that $ \Gamma(R\rightarrow gg) >>\Gamma(R\rightarrow
ggg)$.\footnote{Because of the dominance of two-body phase space when
$m_R$ is not large, not because $\alpha_{QCD}^{eff}$ is small in the
$Rgg$ vertex.}  This just means that where we encountered $\Gamma_R
b_{R \rightarrow gg}$ in our our expression for the absorptive
contribution to the width, we actually mean $\Gamma_R b_{R\rightarrow
{\rm tot gluonic}}$.  With this understanding, that $b_{R\rightarrow
gg}$ is to be identified with $b_{R \rightarrow {\rm tot gluonic}}$,
(\ref{B}) holds even for light $m_R$.

\section{Determination of the Full Amplitude}
\label{sec:fullamp}
\hspace*{2em}
The number of gluons in the intermediate state, as well as the
distinction between real and virtual gluons, depends on the scale size
which has been chosen.  This is analogous to how the identification of
an event at LEP as a 2-, 3-, or 4-jet event depends on the
resolution-size chosen for the jets.  With a low-resolution definition
of a gluon, most of the amplitude for $V \rightarrow \gamma X$ will be
contained in the 2-gluon state, while as the resolution is increased,
states with more gluons will become more important.  Furthermore, a
state which appears under high resolution to contain three real gluons
would appear to have one real and one virtual gluon as the resolution
is lowered and two of the gluons are merged.  While real and
absorptive contributions are separately resolution-size dependent, the
predicted total rate of interest will not be.  It is clearly important
to compute real and imaginary parts consistently, however.

We might try to estimate the real part of the amplitude whose
absorptive part we determined in the previous section by using
dispersion relations, avoiding the use of pQCD except to determine
subtraction constants when needed.  The feasibility of this idea can
be assessed in the regime of applicability of pQCD by trying to
recover the real part of the full pQCD amplitude for
$\Upsilon\rightarrow\gamma\eta_c$ from its absorptive part (see
eq. (\ref{H^PS})) in the physical region, $m_R < m_V$.  Doing so, one
sees that the cut corresponding to $R \rightarrow \gamma V$ is
essential to obtaining the correct result\footnote{GRF thanks
P. Landshoff for useful discussions on this issue.}, so that we cannot
use dispersion relations over experimentally determined quantities to
obtain the full amplitude from the absorptive contribution, and must
find some alternative.

Even when $m_R$ is small, pQCD gives a good approximation for the $V
\rightarrow \gamma g g$ vertex (see Fig. \ref{VR.ps}), due to the
heavy quark in $V$.  The problem when $m_R$ is small and the
resolution scale of the gluons is large, is that pQCD and the
non-relativistic potential does not give a correct description of the
coupling of $R$ to the two gluons.  Perturbatively, the rate for $R
\rightarrow g g$ is, for various $J^{PC}$'s:\footnote{The decay rates
of the S-wave quarkonia are given in Reference~\cite{kuhn_ono} and the
decay rates of the P-wave quarkonia are given in
Reference~\cite{barbieri_gatto_kogerler}.}
\begin{eqnarray}
\Gamma_{0^{-+}}&=&\frac{8}{3}\ \frac{\alpha_{s}^{2}}{m_R^{2}}\ |R_{S}(0)|^{2}
\hspace{100pt}(a)
\nonumber\\
\Gamma_{0^{++}}&=&96\ \frac{\alpha_{s}^2}{m_R^{4}}\ |R'_{P}(0)|^{2}
\hspace{100pt}(b)\nonumber\\
\Gamma_{1^{++}}&\simeq&9.6\ \frac{\alpha_{s}^{2}}{m_R^{4}}\ |R'_{P}(0)|^{2}
\hspace{100pt}(c)\nonumber\\
\Gamma_{2^{++}}&=&\frac{128}{5}\ \frac{\alpha_{s}^{2}}{m_R^{4}}\
        |R'_{P}(0)|^{2}\ .
\hspace{90pt}(d)\nonumber\\
\label{decaywidths}
\end{eqnarray}\\
$R(0)$ is the radial wave function of the bound state at $r=0$ and
$R'(0)$ is its derivative at $r=0$.

Now let us we take the pQCD prediction for $\Gamma(V\rightarrow \gamma
R)$ of refs. \cite{kkks,guberina_kuhn}, and use these expressions
(\ref{decaywidths}) for the width of $R$, to rewrite the pQCD formulae
for $\Gamma(V \rightarrow \gamma R)$ in terms of the width $\Gamma_R$
rather than $R(0)$ or $R'(0)$.  This yields our central result which,
for a $0^{+-}$, is:
\begin{equation}
b(V\rightarrow \gamma R) = b( V \rightarrow \gamma gg) \frac{m_R
\Gamma_R b_{R \rightarrow gg}}{8 \pi (\pi^2 - 9) m_V^2} (1 -
(\frac{m_R}{m_V})^2) |{\hat{\cal H}}^{PS}(x)|^2,
\label{final}
\end{equation}
where $x = (1-(\frac{m_R}{m_V})^2)$ and
\begin{eqnarray}
{\hat{\cal H}}^{PS}(x) = \frac{4}{x} \left\{ L_2(1-2x) - L_2(1) -
\frac{x}{1-2x}ln~2x \right.
\nonumber \\
 \left. - \frac{1-x}{2-x} \left( 2 L_2(1-x) - 2 L_2(1) +
\frac{1}{2} ln^2(1-x) \right) \right\}
\nonumber \\
+ i \pi \frac{4}{x}\frac{1-x}{2-x}ln~(1-x).
\label{H^PS}
\end{eqnarray}
\begin{equation}
L_2(x) = -\int_0^x dt \frac{ln(1-t)}{t}.
\end{equation}
For $0^{++},~1^{++},~2^{++}$ the $\frac{1}{8 \pi}$ factor in (\ref{final})
becomes
$\frac{4}{3 \pi},~\frac{\simeq 40}{3 \pi}, ~\frac{5}{\pi}$ respectively, and
thei
$\hat{{\cal H}}^{PS}$ is replaced by the function for the appropriate
$J^{PC}$ given in \cite{kkks}.  Explicit forms for $\hat{{\cal
H}}^{S,V,T}$ can also be found in \cite{cakir}.

The form of the result (\ref{final}) agrees with that of the unitarity
calculation (\ref{B}) in being proportional to
$\Gamma_R~b_{R\rightarrow gg}$.  Its absorptive part correctly
reproduces the absorptive part obtained by unitarity, using pQCD to
obtain $V\rightarrow \gamma gg$, as we have argued is reliable on
account of the large quark mass in $V$.  By construction it agrees
with the full pQCD non-relativistic potential result, when that is
applicable to computing $\Gamma(R \rightarrow gg)$, i.e., when $R$
contains massive quarks.  Finally, for light mesons it agrees with the
light-cone QCD result of Kuhn\cite{kuhn:lightquarks} when the $f_R$
and $\alpha_s$ dependence is removed in favor of $\Gamma(R \rightarrow
gg)$ computed from (\ref{decaywidths}a) using the relation (\ref{fR})
between $f_R$ and $R(0)$.

We therefore propose that (\ref{final}) is a good approximate
expression for OZI-suppressed radiative production of {\it any}
resonance.  Further work is needed to assign an error to it, because
it involves not only the $O(\alpha_s)$ error due to dropping the hard
three gluon states, but also the assumption that the relative size of
real and imaginary parts is correctly given by pQCD even when the
non-relativistic potential cannot be used to evaluate the $R-gg$
vertex.  A productive line of reasoning to put this on a more rigorous
footing and allow estimation of the error would be to follow Kuhn's
discussion for light $q\bar{q}$ mesons\cite{kuhn:lightquarks}, since
the dynamics of his light $q\bar{q}$ system should be similar to a
glueball and an $\eta_{\tilde{g}}$, if $m(\tilde{g})$ is close enough
to the mass of a strange quark.  Note that our discussion of the
resolution-size dependence of the {\it description} of the
intermediate state (number of gluons, their virtuality...) suggests
that corrections to (\ref{final}) come only from the hard 3-gluon
intermediate state and thus are $O(\alpha_s(M_V/2))$.  As long as data
rather than a model is used for $\Gamma_R$, and one does not attempt to
distinguish between real and imaginary part contributions to the total
rate, soft gluon corrections may be completely included.  A
careful treatment with scale-size introduced is required to
verify this conjecture.

It is interesting that the contribution of the imaginary part of
(\ref{final}) is very tiny compared to that of the real part, but the
total prediction of (\ref{final}) is similar to the unitarity
prediction (\ref{B}), using data for $\frac{d\Gamma_{V \rightarrow \gamma
X}}{ds}$ and a crude model for ${\cal P}^J_{S;L}$\cite{cakir}.
This is consistent with the intuition that typical hadrons are
composed of somewhat virtual quarks and gluons, so that the relative
importance of virtual gluons in the pQCD calculation is much greater
than that of virtual hadrons when a hadronic basis for the calculation
is used.

\section{Some Applications to Data}
\label{data}
\begin{figure}
\epsfxsize=7in
\epsffile{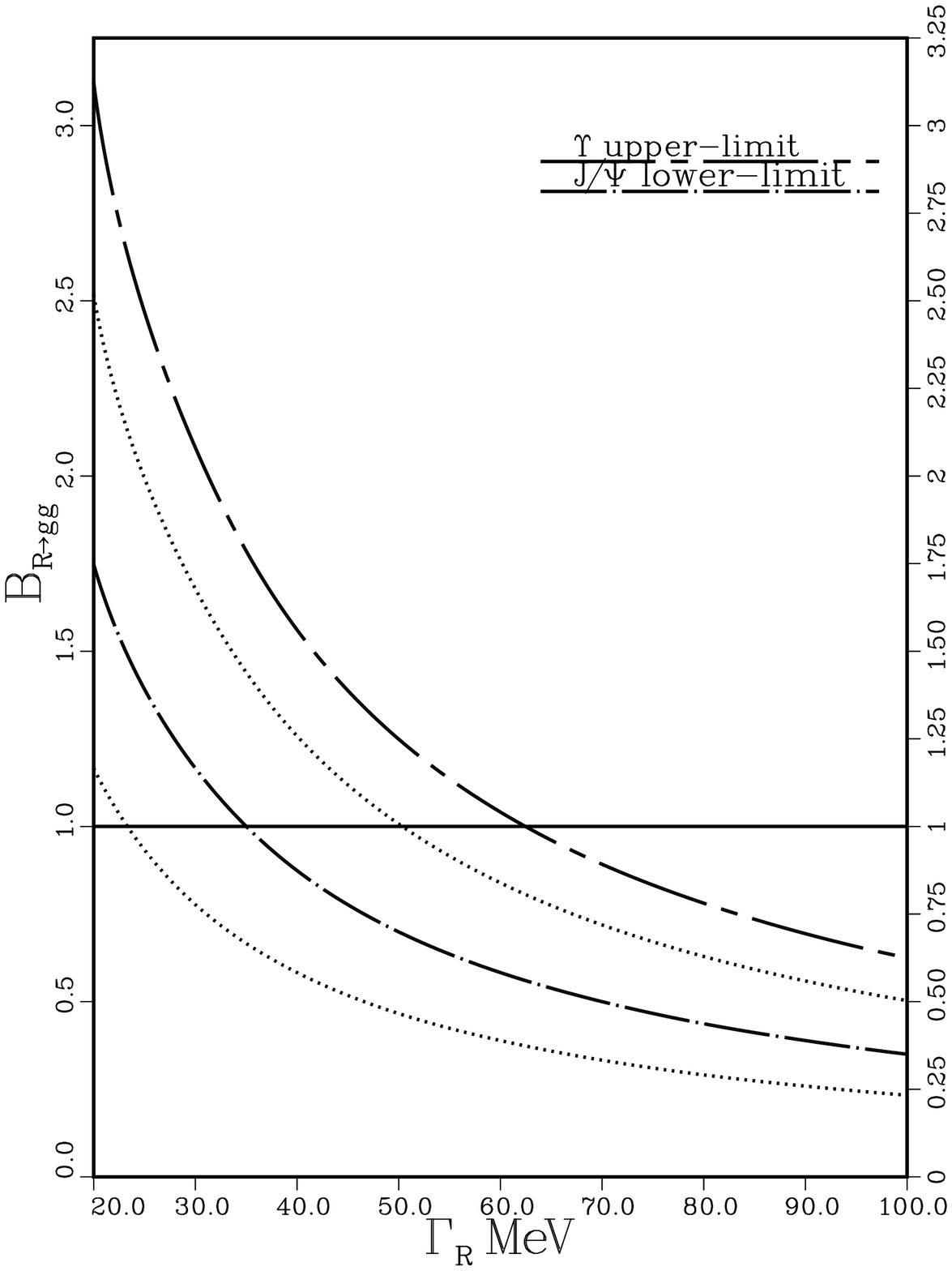}
\caption{The lower limit obtained on the gluonic branching fractions of the
resonance $0^{-+}(1416)$ reported by MARKIII (dot-dashed lines, with
dotted lines at $\pm1$ s.d.) and upper limits from CUSB (long-dashed lines).}
\label{MARK3/1/gg.ps}
\end{figure}

\begin{figure}
\epsfxsize=7in
\epsffile{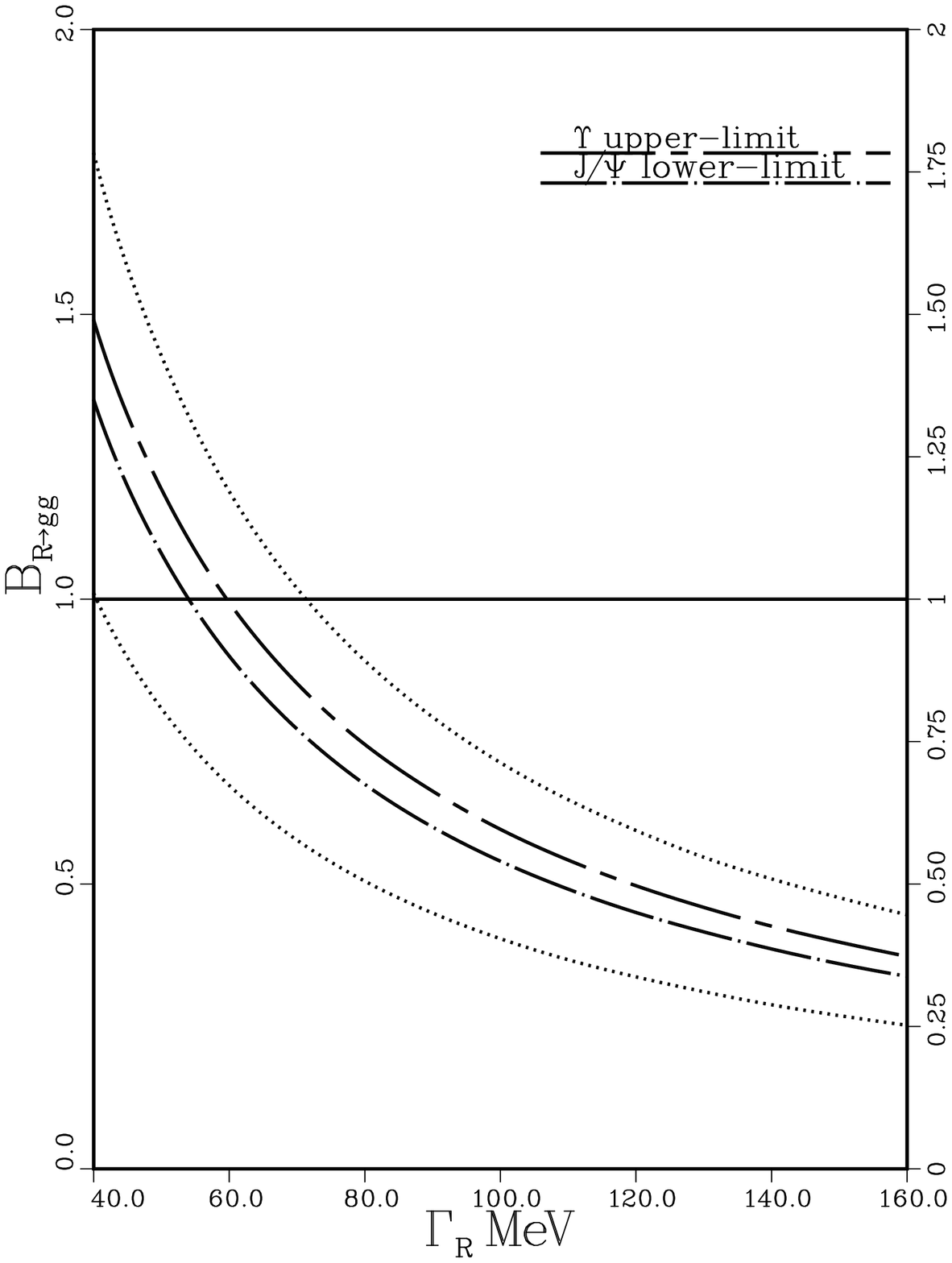}
\caption{The lower limit obtained on the gluonic branching fractions of the
resonance $0^{-+}(1490)$ reported by MARKIII (dot-dashed lines, with
dotted lines at $\pm1$ s.d.) and upper limits from CUSB (long-dashed lines).}
\label{MARK3/2/gg.ps}
\end{figure}

\begin{figure}
\epsfxsize=7in
\epsffile{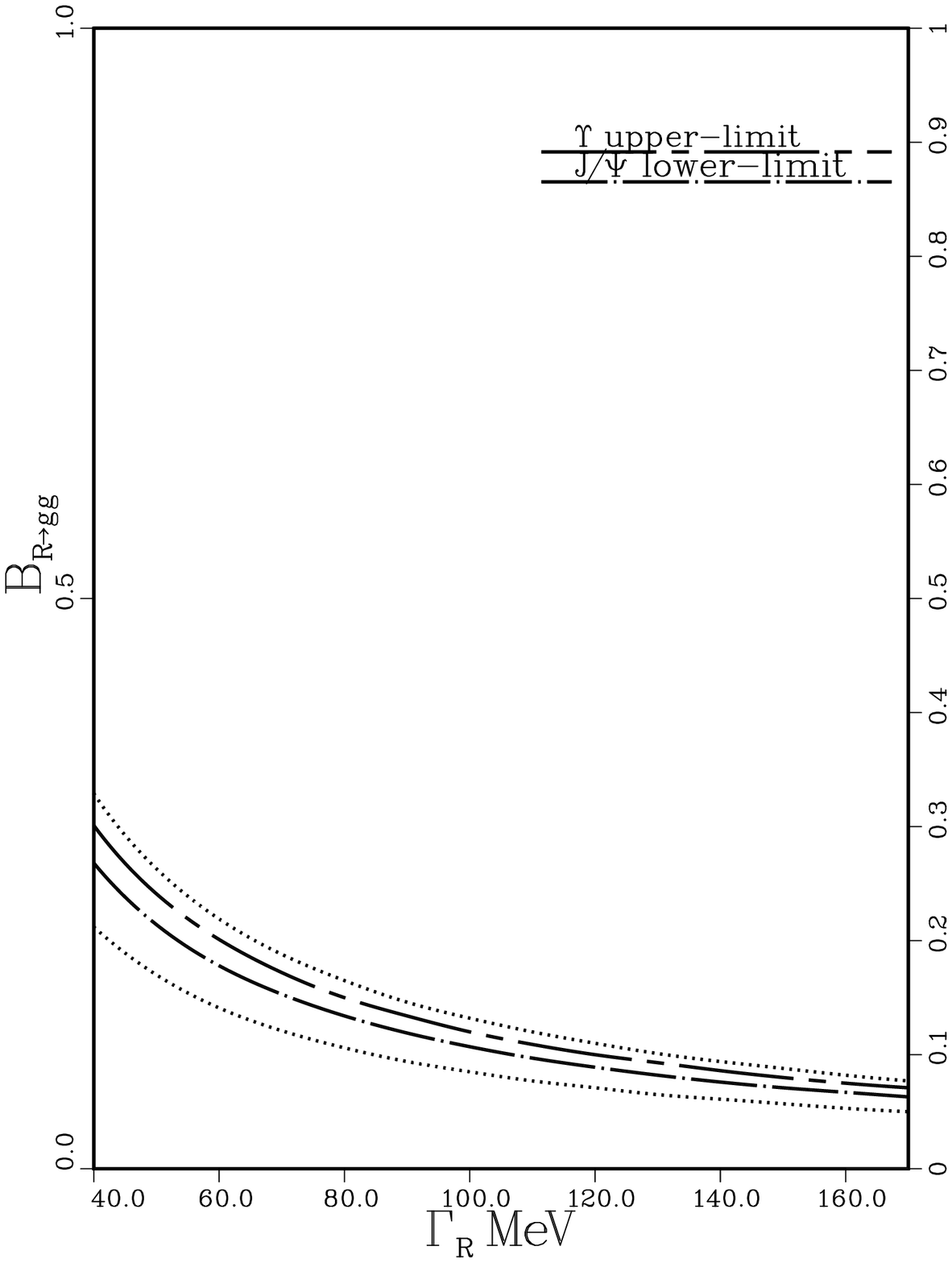}
\caption{The lower limit obtained on the gluonic branching fractions of the
resonance $1^{++}(1443)$ reported by MARKIII (dot-dashed lines, with
dotted lines at $\pm1$ s.d.) and upper limits from CUSB (long-dashed lines).}
\label{MARK3/3/gg.ps}
\end{figure}

\begin{figure}
\epsfxsize=7in
\epsffile{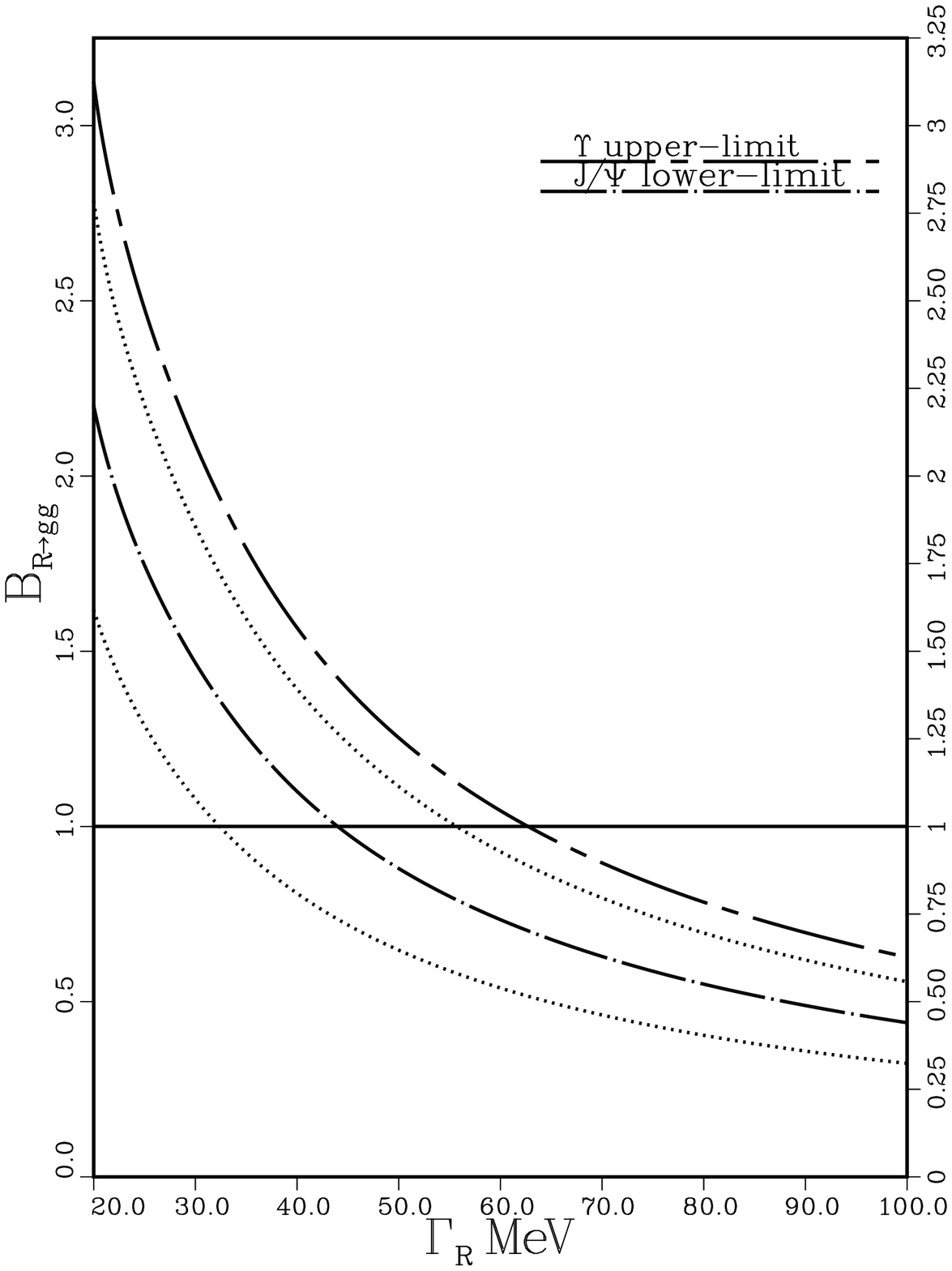}
\caption{The lower limit obtained on the gluonic branching fractions of the
resonance $0^{-+}(1421)$ reported by DM2 (dot-dashed lines, with
dotted lines at $\pm1$ s.d.) and upper limits from CUSB (long-dashed lines).}
\label{DM2/1/gg.ps}
\end{figure}

\begin{figure}
\epsfxsize=7in
\epsffile{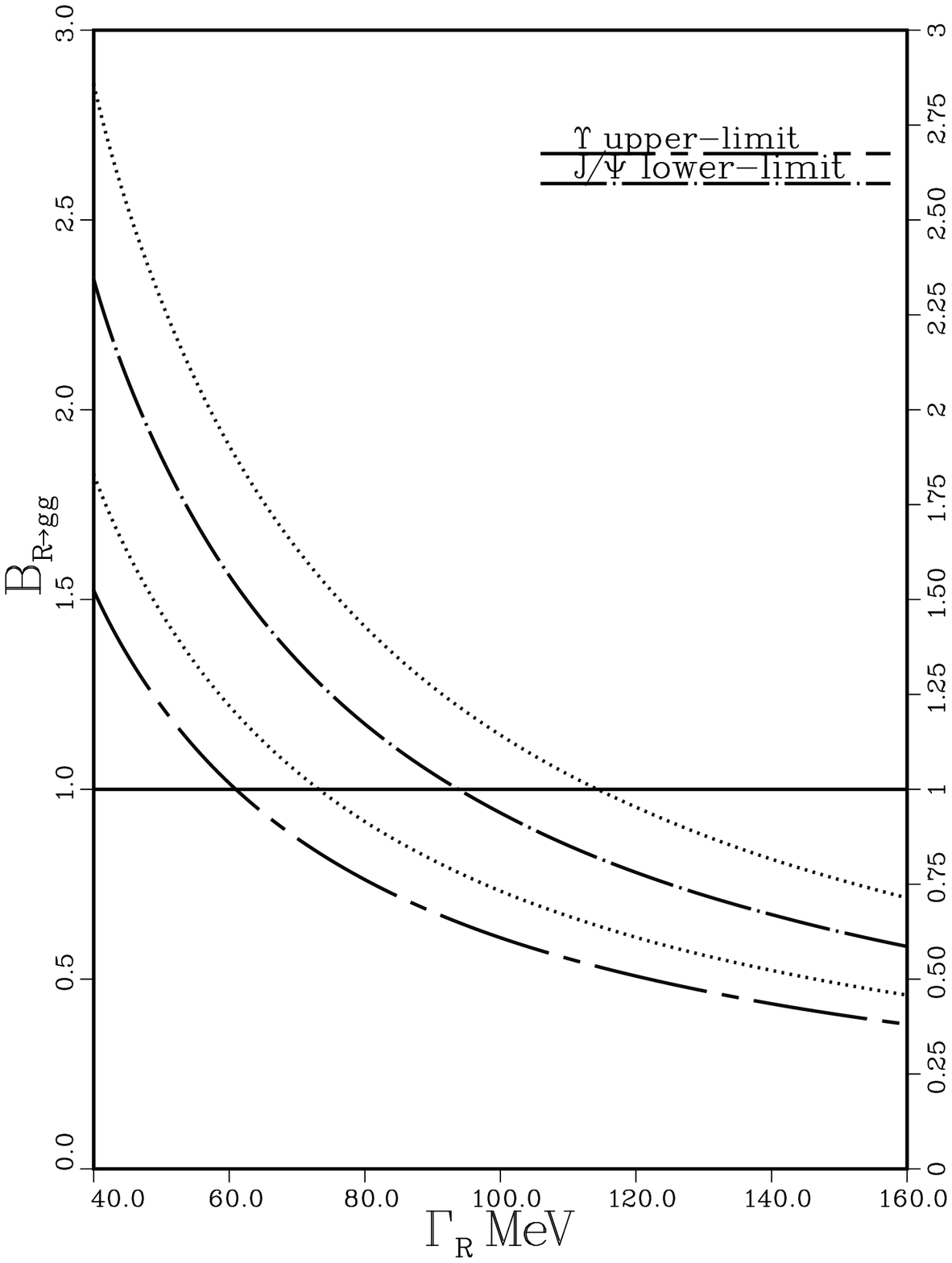}
\caption{The lower limit obtained on the gluonic branching fractions of the
resonance $0^{-+}(1459)$ reported by DM2 (dot-dashed lines, with
dotted lines at $\pm1$ s.d.) and upper limits from CUSB (long-dashed lines).}
\label{DM2/2/gg.ps}
\end{figure}

\begin{figure}
\epsfxsize=7in
\epsffile{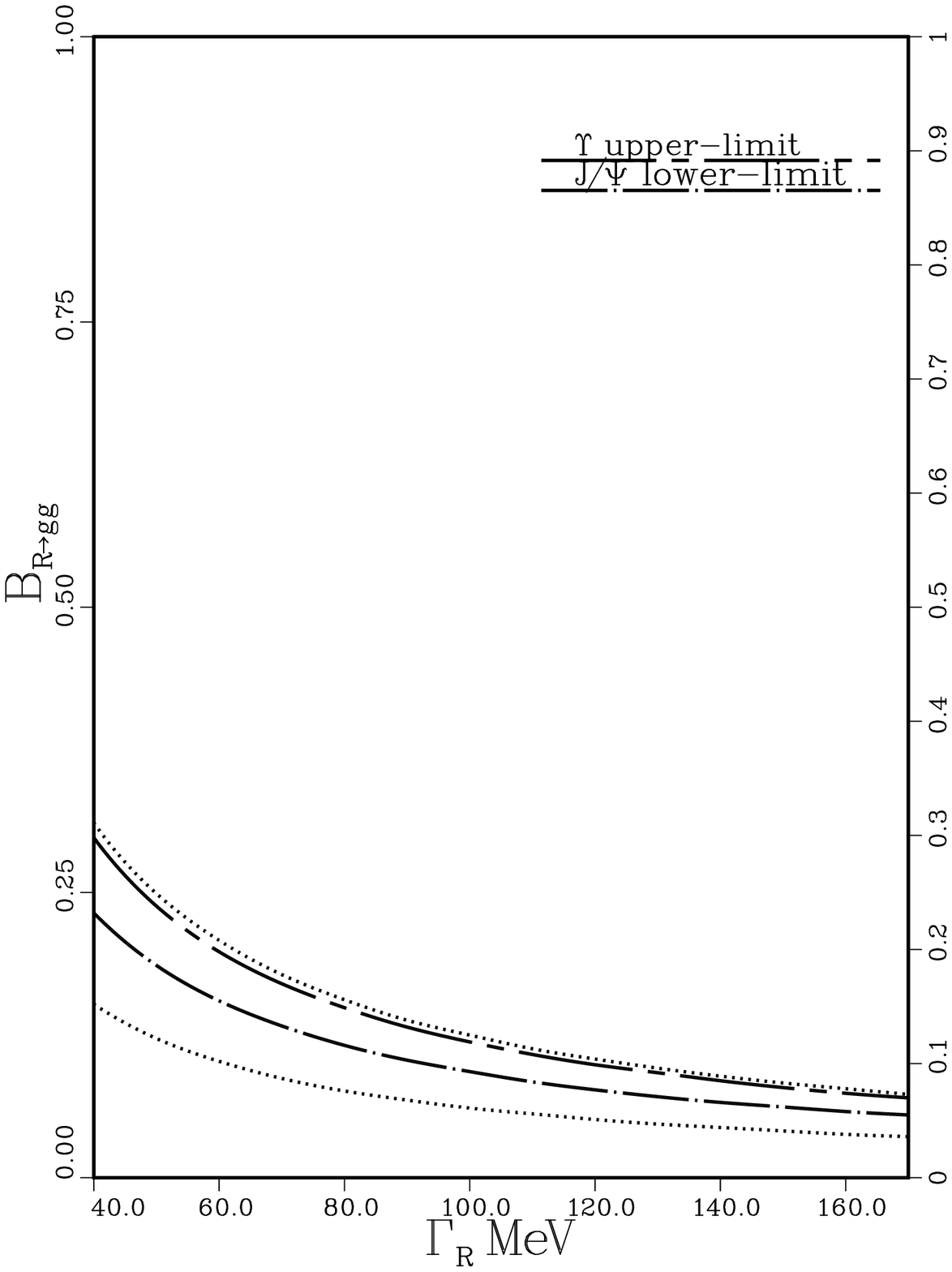}
\caption{The lower limit obtained on the gluonic branching fractions of the
resonance $1^{++}(1462)$ reported by DM2 (dot-dashed lines, with
dotted lines at $\pm1$ s.d.) and upper limits from CUSB (long-dashed lines).}
\label{DM2/3/gg.ps}
\end{figure}

\begin{figure}
\epsfxsize=7in
\epsffile{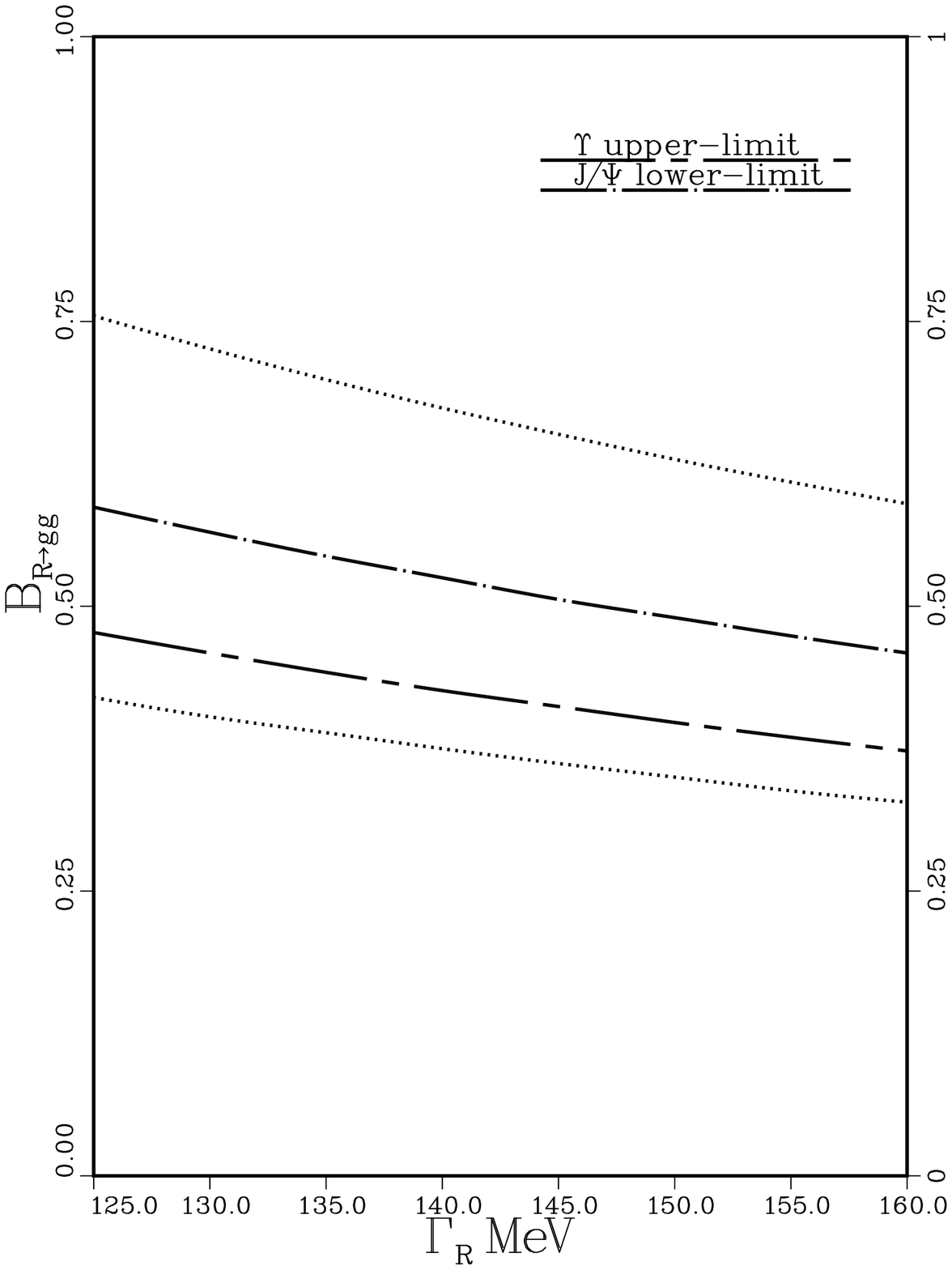}
\caption{CUSB upper limit (long-dashed lines) on $b(\eta(1490) \rightarrow
gg)$, and lower limits from $\Gamma(J/\Psi \rightarrow \gamma \eta(1490)
\rightarrow \gamma X) = 1.0 \pm 0.2 \times 10^{-3}$ (dot-dashed lines, with
dotted lines at $\pm1$ s.d.).}
\label{eta1490.ps}
\end{figure}

\begin{figure}
\epsfxsize=7in
\epsffile{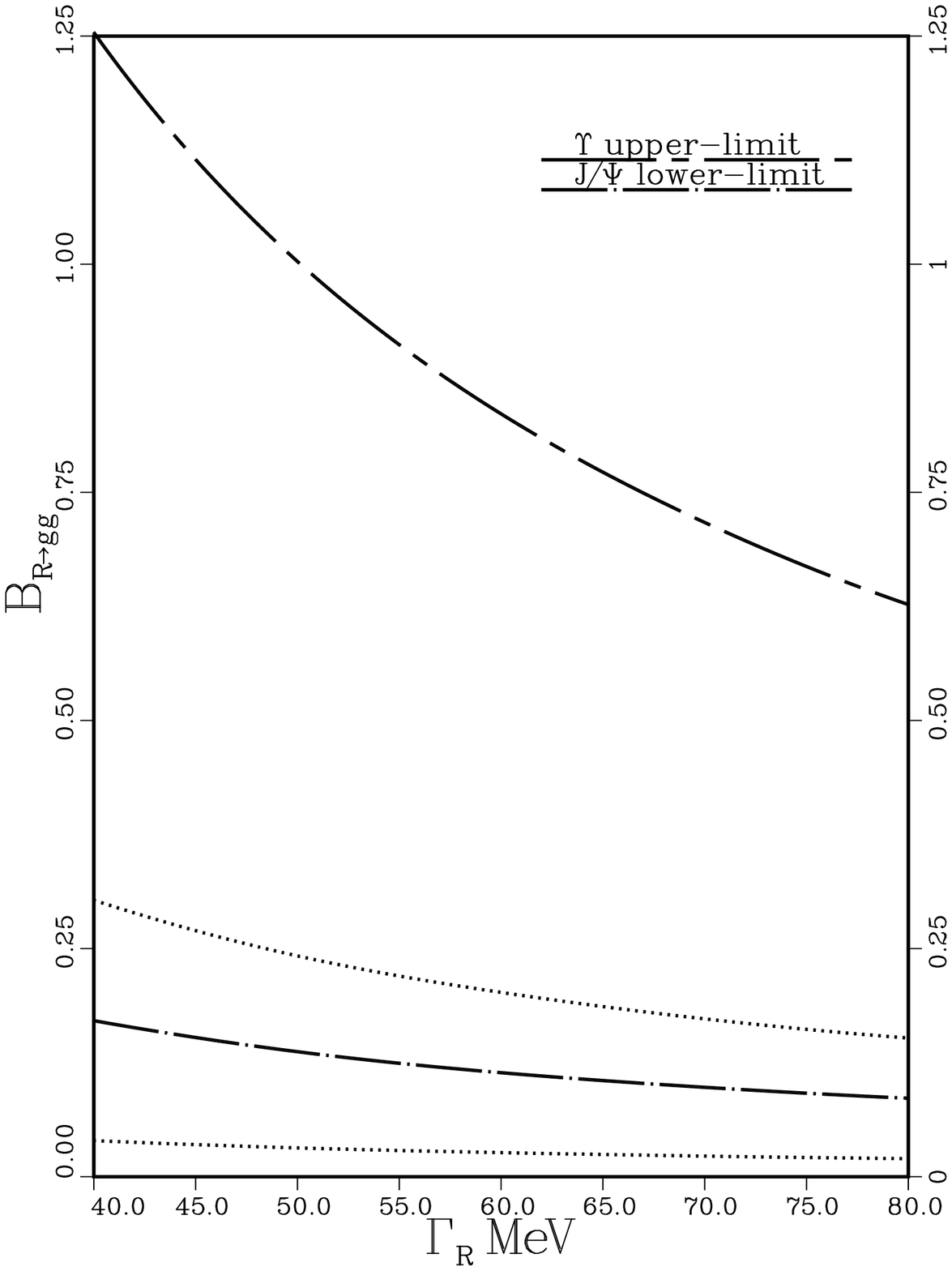}
\caption{CUSB upper limit (long-dashed lines) on $b(\eta(1760) \rightarrow
gg)$, and lower limits from $\Gamma(J/\Psi \rightarrow \gamma \eta(1760)
\rightarrow \gamma X) = 1.3 \pm 0.9 \times 10^{-4}$ (dot-dashed lines, with
dotted lines at $\pm1$ s.d.).}
\label{eta1760.ps}
\end{figure}

\begin{figure}
\epsfxsize=7in
\epsffile{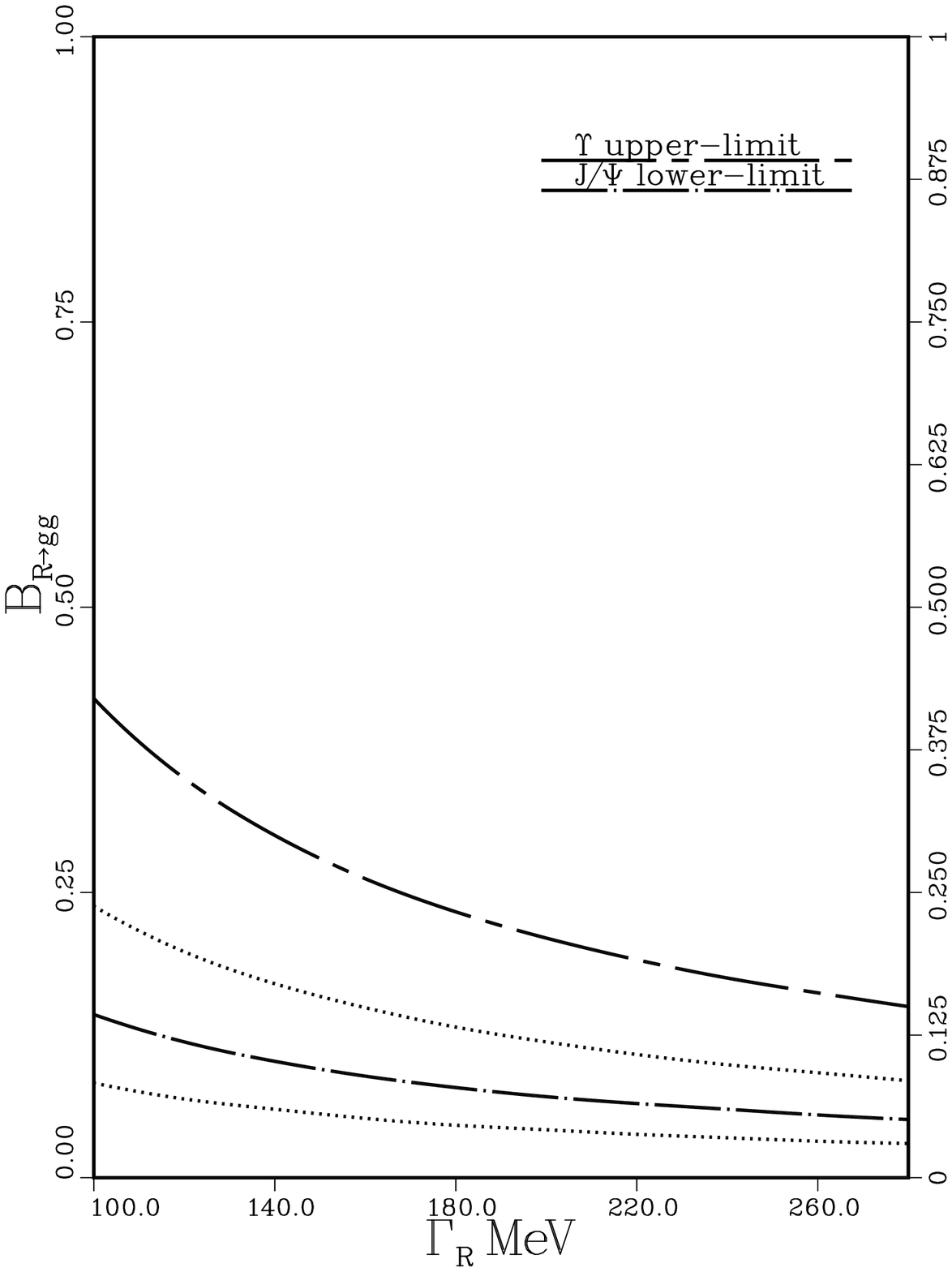}
\caption{CUSB upper limit (long-dashed lines) on $b(\eta(2100) \rightarrow
gg)$, and lower limits from $\Gamma(J/\Psi \rightarrow \gamma \eta(2100)
\rightarrow \gamma X) = 2.9 \pm 0.6 \times 10^{-4}$ (dot-dashed lines, with
dotted lines at $\pm1$ s.d.).}
\label{eta2100.ps}
\end{figure}

\begin{figure}
\epsfxsize=7in
\epsffile{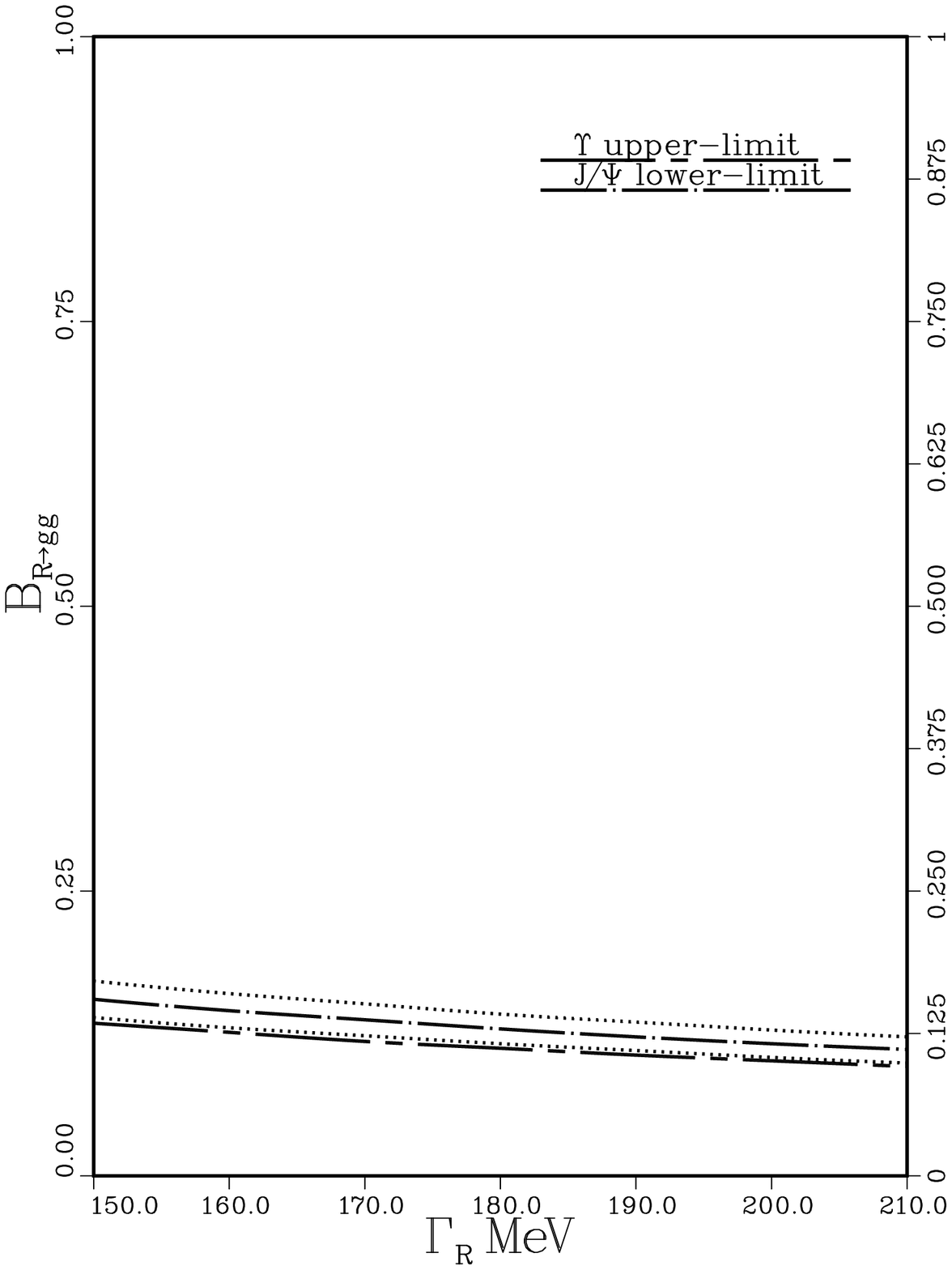}
\caption{CUSB upper limit (long-dashed lines) on $b(f_{2^{++}}(1270)
\rightarrow
gg)$, and lower limits from $\Gamma(J/\Psi \rightarrow \gamma f_{2^{++}}(1270)
\rightarrow \gamma X) = 1.4 \pm 0.2 \times 10^{-3}$ (dot-dashed lines, with
dotted lines at $\pm1$ s.d.).}
\label{f1270.ps}
\end{figure}

\begin{figure}
\epsfxsize=7in
\epsffile{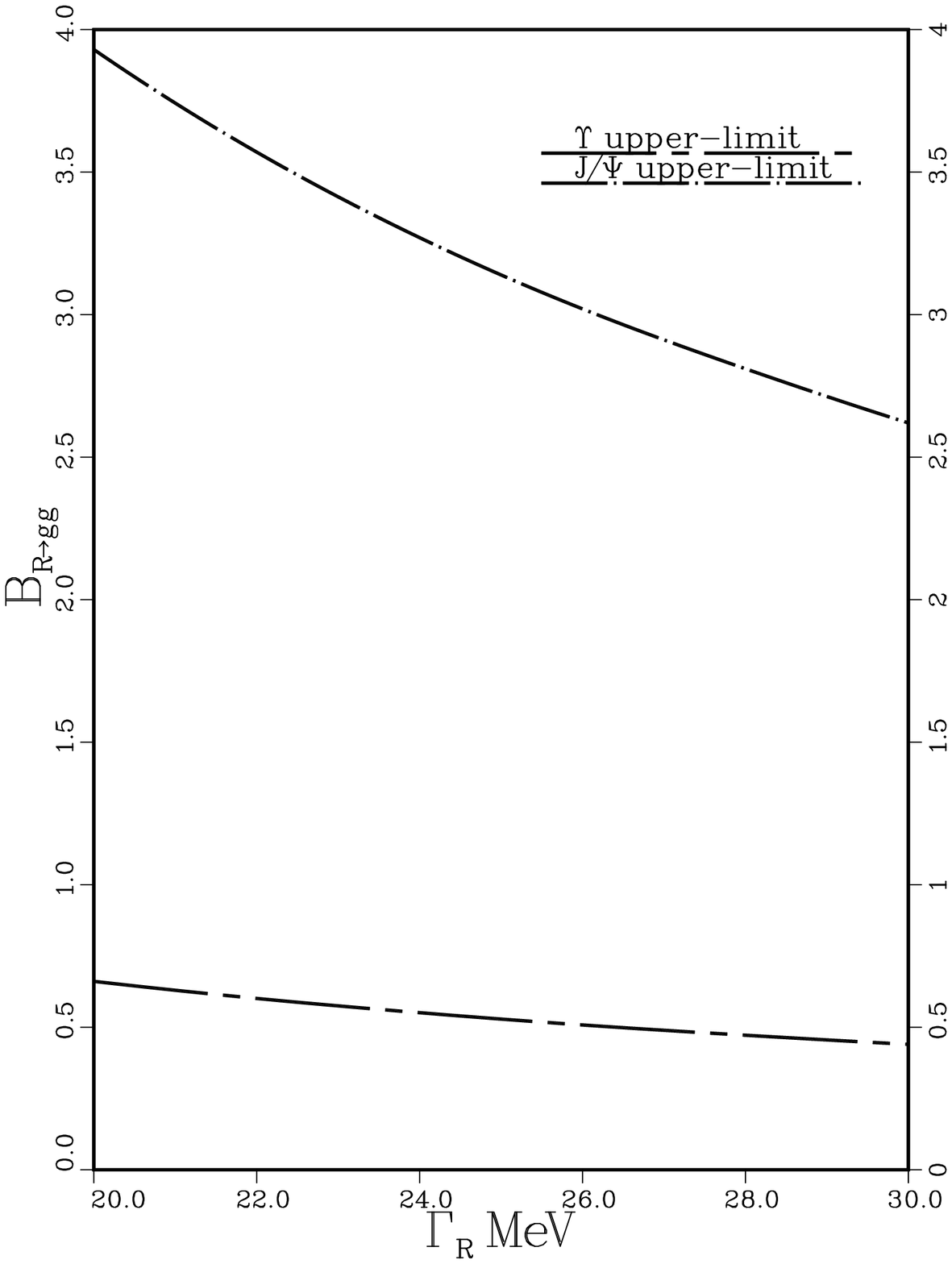}
\caption{CUSB upper limit (long-dashed lines) on $b(f_{1^{++}}(1285)
\rightarrow
gg)$, and lower limits from $\Gamma(J/\Psi \rightarrow \gamma f_{1^{++}}(1285)
\rightarrow \gamma X) = 7.0 \pm 2.0 \times 10^{-4}$ (dot-dashed lines, with
dotted lines at $\pm1$ s.d.).}
\label{f1285.ps}
\end{figure}

\hspace*{2em}
Figs. \ref{MARK3/1/gg.ps}-\ref{f1285.ps} show the upper and lower limits on
$b_{R \rightarrow gg}$ which are obtained by comparing the prediction of
eq. (\ref{final}) with the CUSB and exclusive data.  We take $b(\Upsilon
\rightarrow \gamma X) = 0.03$\cite{upsrad} and $b(J/\Psi \rightarrow \gamma
X) = 0.06$\cite{psirad}.  In general, the branching fraction of the
resonance into the specific mode in which it is observed in the exclusive
radiative decay experiments is not known, although of course it is no
greater than 1.  Except when the branching fraction is known, the exclusive
data therefore only gives a lower limit on $b_{R \rightarrow gg}$.  In some
instances it is seen that the CUSB upper limit and the exclusive lower
limits are quite near, resulting in an estimate of $b_{R \rightarrow g g}$
and the predictions that a) with a modest increase in sensitivity this
resonance should show up in $\Upsilon \rightarrow \gamma X$ and b) the
exclusive branching fraction must be near 1.

The predictions in the figures have some imprecision because the
widths are poorly known in many cases.  Moreover there is some
intrinsic error in our method which we have not estimated, but is at
least $O(\alpha_s)$ from neglect of the 3-hard gluon state.
Nonetheless, the results are interesting and generally plausible.
Even though DM2 and MarkIII disagree on the order and exact widths of
the three resonances they see in the 1440 MeV region, both
pseudoscalars seen in both experiments are likely to be glueballs or
the result of mixing a glueball with a $q\bar{q}$ state.  Details of
these states are given in Table 1. The remaining figures refer to
particles whose radiative production in $J/\Psi$ decay is given in the
PDG datatables (1990 edition, generally).  The $\eta(1490)$ and
$f_{0^{++}}(1720)$ are hard to classify as pure $q\bar{q}$, since
their $b_{R \rightarrow gg}$'s are $\sim 50\%$.  The $\eta(1760)$
could be either a $q\bar{q}$ or gluonic state, given the spread in the
lower and upper limits.  The exclusive production for $f_1(1285)$ is
above the upper limit, given a $12\%$ branching fraction to $K \bar{K}
\pi$, so that there is some internal inconsistency.  Perhaps the
problem is in the CUSB upper limit: they have a poorly-understood
contribution to their data from non-resonant processes in the region
$m_R \lessim 1.5$ GeV\cite{tutsmunich}.  It could be wise to mentally
attribute some additional systematic error to the CUSB limits for the
low mass region.  Instead, this discrepancy between upper and lower
limits for the $f_1(1285)$ may reflect the intrinsic error of our
method.  If the latter is the case, our method will not be very useful
unless the problem is isolated to $1^{++}$ production.  The question
can be decided experimentally when the branching ratios
$B(R\rightarrow K \bar{K} \pi)$ are known, and resonances are actually
seen in $\Upsilon \rightarrow \gamma R$.

Our conclusions above regarding which resonances may be glueballs are
generally consistent with other indicators.  One interesting point which we
do not pursue here is the possibility that there is actually an ``extra''
flavor singlet pseudoscalar around 1.4 GeV\cite{chanowitz,chung}, compared
to expectations from filling known $q \bar{q}$ nonets and the predicted
glueball spectrum.  This will have to await further experimental
elucidation of this mass region.  Other applications of this method,
including more figures and tables, can be found in \cite{cakir}.

\section{Limits on Gluino Mass}
\label{gluino}
\hspace*{2em}
CUSB has claimed\cite{tutsmunich} to exclude a gluino with mass in the
range $0.21 < m(\tilde{g}) < 3.6$ GeV, without however seeing any
other resonances in their spectrum.  It is on account of the strong
gluonic coupling of the gluino compared to the quark that it was
noted\cite{keung_khare,kuhn_ono,goldman_haber} that, when pQCD and the
non-relativistic potential model are applicable, the
$\eta_{\tilde{g}}$ will be produced at a significantly larger level
than a $Q \bar{Q}$ resonance, and in particular at a level which
should be seen at CUSB.  Let us now address the question as to
the range of validity of these calculations which rely on pQCD and the
non-relativistic potential
model\footnote{Refs. \cite{keung_khare,goldman_haber} themselves
remark that their analysis applies only for $m(\eta_{\tilde{g}}) \gtrsim
3$ GeV.}, and find out what can be said when we cannot use them.

{}From eq. (\ref{final}) and the bound (\ref{cusb_bound}) on $b(\Upsilon
\rightarrow \gamma X)$, knowing the radiative branching fraction of
the $\Upsilon$ to be 0.03, we can extract an upper bound on the width
of the $\eta_{\tilde{g}}$ as a function of its mass.  This is shown as
the solid curve in Fig. \ref{widthlim}.  Evidently, if the
$\eta_{\tilde{g}}$ has a width less than $\sim 40$ MeV, it cannot be
excluded for any mass.  Fig. \ref{widthlim} also shows the
non-relativistic potential model/pQCD (nrpm/pQCD) prediction for the
width, for $\Lambda_{QCD} = $ 100 and 200 MeV (the lower and upper
dashed curves,
respectively).  It is
obtained\cite{keung_khare,kuhn_ono,goldman_haber} by replacing the
$\frac{8}{3}$ in eq. (\ref{decaywidths}a) by $18$, and using $|R(0)|^2
= (\frac{9 m_R}{4 m(\eta_c)})^{\frac{3}{2}} |R(\eta_c)(0)|^2$.

Based on the fact that the experimental limit
lies above the theoretical prediction for $m(\eta_{\tilde{g}}) \lessim 7$
GeV, CUSB concluded that they could exclude gluino masses below 3.6
GeV\cite{tutsmunich}\footnote{Aside from being insensitive if the
$\eta_{\tilde{g}}$ is too light to decay to pions, which they say
occurs for a gluino mass less than 0.21 GeV.}.  However for
$m(\eta_{\tilde{g}}) \lessim 3$ GeV the predicted width becomes very
large and may signal a failure of this model for the width.
Physically, the way that the width of an $\eta_{\tilde{g}}$ can be
large is for the constituents to be so massive that the bound state is
very small, leading to a large wavefunction at the origin, while due
to the large color charge compared to a quark, the intrinsic
gluino-gluon coupling strength is big compared to the quark-gluon
coupling.  Since the color charge is important both in making the
system tightly bound and thus concentrated at the origin, and in
increasing the coupling to the final state gluons, the effect of a
larger color charge enters twice as we saw above, leading to the
observation of refs. \cite{keung_khare,kuhn_ono,goldman_haber} that an
$\eta_{\tilde{g}}$ would be very prominent in the radiativge decay
spectrum if the non-relativistic potential and pQCD were relevant.

However if the gluino is lighter than perhaps $\frac{1}{2} - 1$ GeV,
the bound state properties should resemble those of gluons and strange
quarks.  We do not know much about the former, but we do know that the
latter form hadrons whose decay constants are remarkabaly similar.
(See fig. \ref{ffac.ps} which shows the measured pseudoscalar decay
constants.)  These which are related to the wavefunction at the origin
by\footnote{This formula can be derived or found in
refs. \cite{fR_R(0)1,fR_R(0)2} and also is implicit in the connection
between refs. \cite{kuhn:lightquarks} and \cite{kkks}.}
\begin{equation}
f_R = \sqrt{\frac{3 R_R^2(0)}{\pi m_R}}\hspace{5mm}.
\label{fR}
\end{equation}
The similarity in value of the various nonet pseudoscalar decay
constants means that for bound states of light constituents
$|R(0)|^2\sim m_R ({\rm hadronic volume})^{-1}$.  The volume of the
hadron is mainly governed by the confinement scale and has little to
do with the mass of the light constituent.  Furthermore, with light
constituents, the individuality of the constituent is lost amidst the
sea of gluons and $q\bar{q}$ pairs, so that whether the constituents
are quarks or gluinos should not matter much, if the gluino is light.
If this is a correct interpretation, one would expect $f_R \sim
f_{\pi}$ for $R$ a pseudoscalar bound state of light gluinos.  Now
taking $R(0)$ for the $\eta_{\tilde{g}}$ from eq. (\ref{fR}) with $f_R
= $120 MeV and using eq. (\ref{decaywidths}a), one obtains the
$\eta_{\tilde{g}}$ width shown in the dot-dashed curves (``mesonic
wavefunction model'') in fig. \ref{widthlim}.  The upper curve
corresponds to the nrpm/pQCD expression (\ref{decaywidths}a),
replacing $\frac{8}{3}$ in (\ref{decaywidths}) by 18, with $\alpha_s =
\frac{12 \pi}{25} ln (\frac{m_R}{\Lambda_{QCD}})^2$ for $\Lambda_{QCD}
= $100 MeV, and the lower one is obtained by replacing the factor $18
\alpha_s^2$ with 1, suitable if the interaction strength saturates at
low energy scale.  Either way, one obtains a much smaller prediction
for the $\eta_{\tilde{g}}$ width, $O(10$ MeV), than from the
nrpm/pQCD, basically because the nrpm formula for $|R(0)|$ grossly
overestimates it for relativistic constituents.  Note however that
when the constituents are relativistic, the decay $R \rightarrow gg$
presumably probes a larger spatial portion of the wavefunction than
merely the point at the origin, so that the nrpm/pQCD formula itself
(eq. (\ref{decaywidths})) may not be applicable.  Hence using it with
even a perfect estimate for $R(0)$ does not necessarily give a
reliable estimate of the actual width and the curves ``mesonic
wavefunction model'' only serve to give an indication of the large
uncertainty in modeling the width.

Predicting the width of an $\eta_{\tilde{g}}$ is a good problem for
lattice gauge theory.  Until such predictions are available, the most
conservative approach is to take the $\eta_{\tilde{g}}$ width
to be $\frac{1}{10} - 1$ times the typical width of glueball
candidates.  The motivation for this is that in order to communicate with
quarks, the gluino-pair must first convert to gluons, requiring at least
two powers of $\alpha_s$ more than are present in a glueball decay rate.
At the same time, since the system is strongly interacting these factors of
$\alpha_s$ need not be small, leading to the above estimate.

Thus we argue that since CUSB does not see glueballs, its data cannot
be used to exclude an $\eta_{\tilde{g}}$ of a similar mass.  For a
resonance of $\sim 1.5$ GeV, the width limit from CUSB is $\sim 70$
MeV, roughly the width of the pseudoscalar glueball candidates in the
iota region (see Table 1), and we cannot exclude an $\eta_{\tilde{g}}$
in this region.  The mass range between this and $\sim3$ GeV is
ambiguous.  At its upper end, even though we cannot have complete
confidence in the nrpm/pQCD calculation on theoretical
grounds\footnote{We cannot be completely confident due to the large
color charge of gluinos: the factor $\frac{8}{3} \alpha_s^2$ for $q
\bar{q}$ becomes $18\alpha_s^2$ for gluinos, so that perturbation
theory is less reliable than for an $\eta_c$ of the same mass.}, the
CUSB limit of $\sim 50$ MeV is well below the nrpm/pQCD prediction of
$150 - 250$ MeV, so there is a comfortable margin of error.  At the
lower end of the range, glueballs exist and they certainly must be
visible in the experiment before drawing conclusions about the absence
of an $\eta_{\tilde{g}}$.  When the width of an $\eta_{\tilde{g}}$ has
been well determined from lattice QCD, one can learn from
eq. (\ref{final}) how sensitive a search is required to observe them
in $V \rightarrow \gamma R$.  Until that time, to be conservative we
conclude that the CUSB experiment can only be used to rule out the
range $ \sim 3 \lessim m(\eta_{\tilde{g}})
\lessim 7 $ GeV.

\section{Summary}
\label{summary}
\hspace*{2em}
We have proposed a method of predicting the branching ratio for production
of any resonance in $V \rightarrow \gamma R$, which only requires knowing
the mass, total width and gluonic branching fraction $b_{R \rightarrow gg}$
of the resonance.  We applied it to determining or obtaining limits on
$b_{R \rightarrow gg}$ for a number of known flavor singlet resonances,
identifying the best glueball candidates as the ones for which $b_{R
\rightarrow gg}$ can be near 1.  Of the states we examined, these are the
two pseudoscalars in the 1440 MeV region, the $f_{0^{++}}$, and possibly
the $\eta(1760)$.

We also found limits on the total width of a possible gluino-gluino bound
state, since for such a state one would have $b_{R \rightarrow gg} \sim
1$.  These limits are in conflict with the non-relativistic potential
model/pQCD prediction of refs. \cite{keung_khare,kuhn_ono,goldman_haber},
so that in its region of validity the existance of an $\eta_{\tilde{g}}$
can be excluded, i.e., for the range $3 \lessim m(\eta_{\tilde{g}}) \lessim 7$
GeV.  We argued that for lower masses the nrpm/pQCD calculation is not
applicable, and instead one can only say that the width of the
$\eta_{\tilde{g}}$ is less than the bound shown in Fig. \ref{widthlim},
i.e.,$\sim 70$ MeV for a mass of about 1.5 GeV or $\sim 50$ MeV for a mass
greater
than 2 GeV.


\clearpage

\begin{table}[tp]
\vspace{20pt}
\vbox{\hfill
$
\begin{tabular}{|c|c|c|c|c|}   \hline\hline
$R$ &$m_{R}(MeV)$ &$\Gamma_{R}(MeV)$&$B^{exp} \mbox{\small$(J/\Psi \rightarrow
 \gamma R,R\rightarrow K\bar{K} \pi)$} $
                &$B^{pred}/b_R$ \\      \hline\hline
$0^{-+}_{(1426)}$&$1426^{+10.6}_{-9.4}$&$54^{+39.2}_{-31.9}$
        &$0.66^{+0.17+0.24}_{-0.16-0.15} \times 10^{-3}$
        &$1.02^{+0.74}_{-0.60} \times 10^{-3}$\\ \hline
$0^{-+}_{(1490)}$&$1490^{+16.3}_{-17.9}$&$91^{+68.7}_{-49.0}$
        &$1.03^{+0.21+0.26}_{-0.18-0.19}\times 10^{-3}$
        &$1.74^{+1.31}_{-0.94}\times 10^{-3}$\\  \hline
$1^{++}_{(1443)}$&$1443^{+7.6}_{-6.3}$&$68^{+30.1}_{-20.1}$
        &$0.87^{+0.14+0.14}_{-0.14-0.11} \times 10^{-3}$
        &$5.53^{+3.89}_{-1.64} \times 10^{-3}$\\ \hline\hline\hline
$0^{-+}_{(1421)}$&$1421\pm 14$&$63\pm 18$
        &$0.83 \pm 0.13 \pm 0.18 \times 10^{-3}$
        &$1.19\pm 0.34 \times 10^{-3}$\\ \hline
$0^{-+}_{(1459)}$&$1459 \pm 5$&$75 \pm 9$
        &$1.78 \pm 0.21 \pm0.33  \times 10^{-3}$
        &$1.43\pm 0.18\times 10^{-3}$\\  \hline
$1^{++}_{(1462)}$&$1462 \pm 20$&$129 \pm 41$
        &$0.76 \pm 0.15 \pm 0.21 \times 10^{-3}$
        &$10.6\pm 3.43 \times 10^{-3}$\\ \hline\hline
\end{tabular}
$
\hfill}
\vspace{20pt}
\caption{Predicted branching ratios for
$J/\Psi \rightarrow \gamma R,R \rightarrow K\bar{K}\pi$, without the factor
$b_{R}\equiv b_{R \rightarrow gg} \times b(R \rightarrow K \bar{K}
\pi)$, for the three resonances in the $\iota(1430)$ region found by
MarkIII and DM2, respectively.  Gluonic states would have $b_{R
\rightarrow gg} \sim 1$, so that dividing the experimental result by the
last column would produce $b(R \rightarrow K \bar{K} \pi)$ }
\vspace{20pt}
\end{table}

\begin{figure}
\epsfxsize=7in
\epsffile{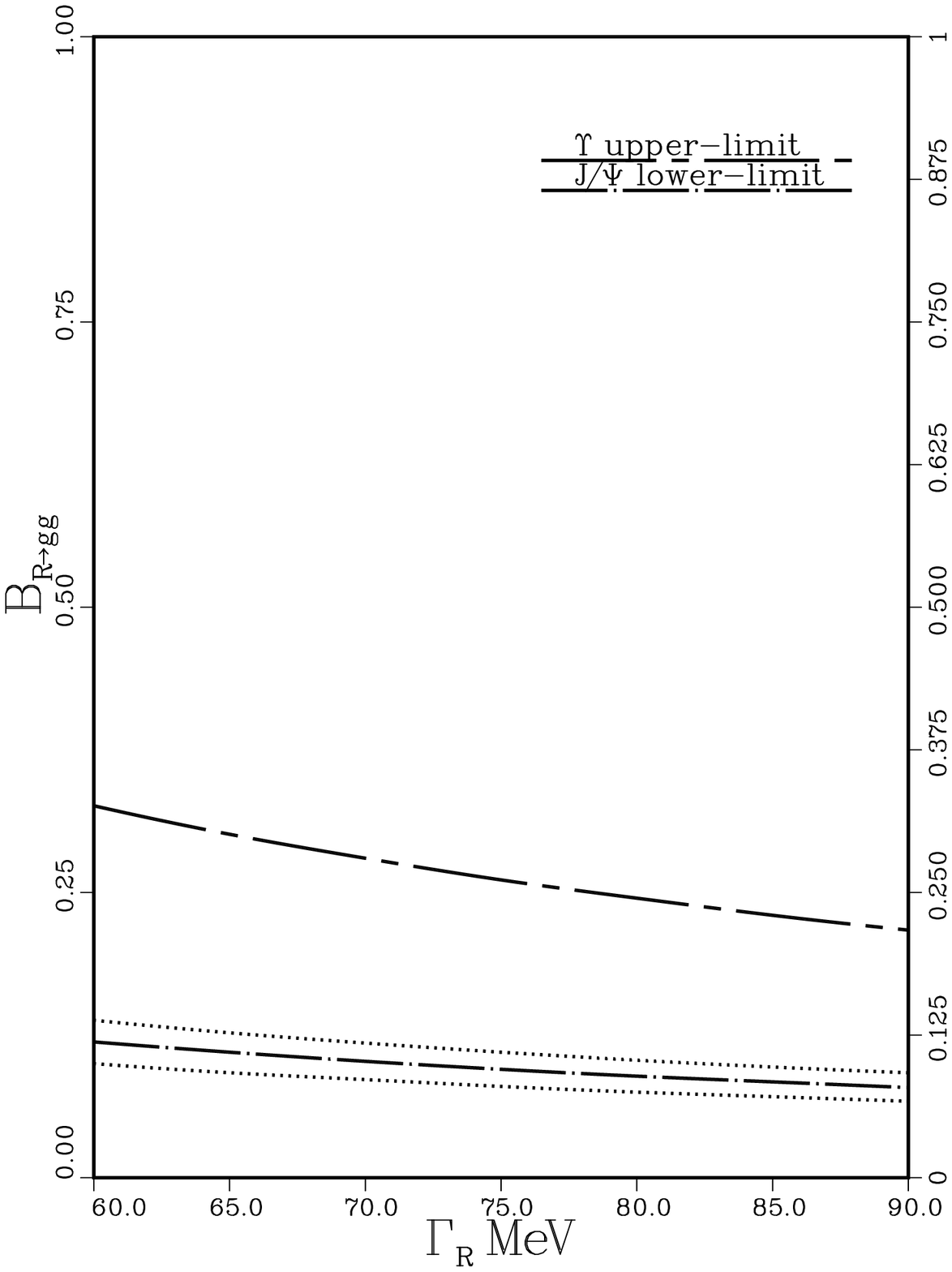}
\caption{CUSB upper limit (long-dashed lines) on $b(f_{2^{++}}(1525)
\rightarrow
gg)$, and lower limits from $\Gamma(J/\Psi \rightarrow \gamma f_{2^{++}}(1525)
\rightarrow \gamma X) = 6.3 \pm 1.0 \times 10^{-4}$ (dot-dashed lines, with
dotted lines at $\pm1$ s.d.).}
\label{f1525.ps}
\end{figure}

\begin{figure}
\epsfxsize=7in
\epsffile{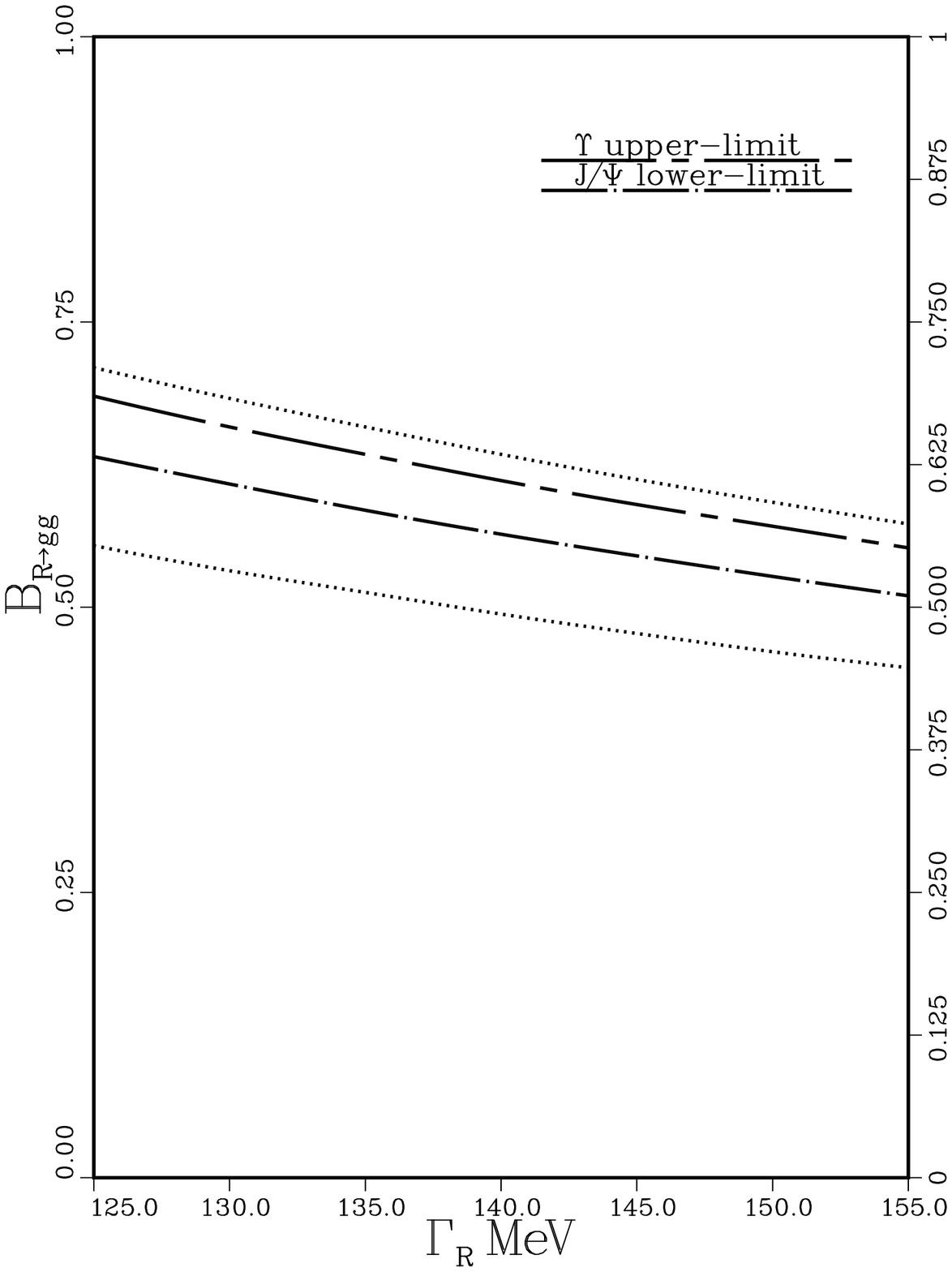}
\caption{CUSB upper limit (long-dashed lines) on $b(f_{0^{++}}(1720)
\rightarrow
gg)$, and lower limits from $\Gamma(J/\Psi \rightarrow \gamma f_{0^{++}}(1720)
\rightarrow \gamma X) = 9.7 \pm 1.2 \times 10^{-4}$ (dot-dashed lines, with
dotted lines at $\pm1$ s.d.).}
\label{f1720s.ps}
\end{figure}

\begin{figure}
\epsfxsize=\hsize
\epsffile{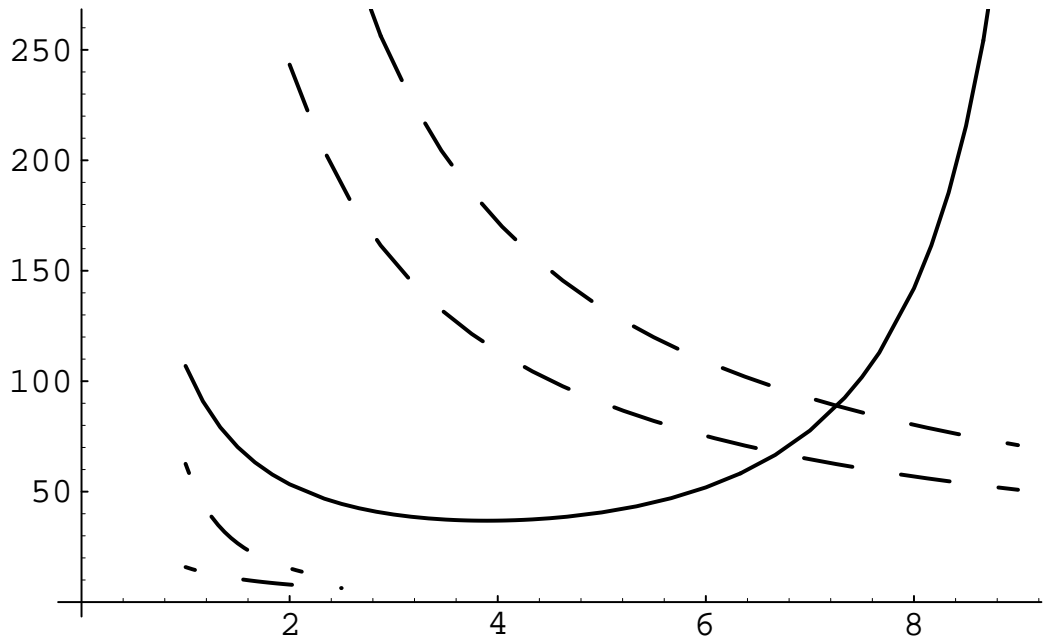}
\caption{Experimental limit on $b_{R \rightarrow gg} \Gamma_R$ for any
resonance produced in $\Upsilon \rightarrow \gamma X$ from CUSB (solid
curve).  Non-relativistic potential + pQCD prediction for width of
an $\eta_{\tilde{g}}$ for $\Lambda_{QCD} = 100$ and 200 MeV (lower and
upper dashed lines).  ``Mesonic wavefunction model'' (see text) for the
width of an $\eta_{\tilde{g}}$ (dot-dashed lines).}
\label{widthlim}
\end{figure}

\begin{figure}
\epsfxsize=8in
\epsffile{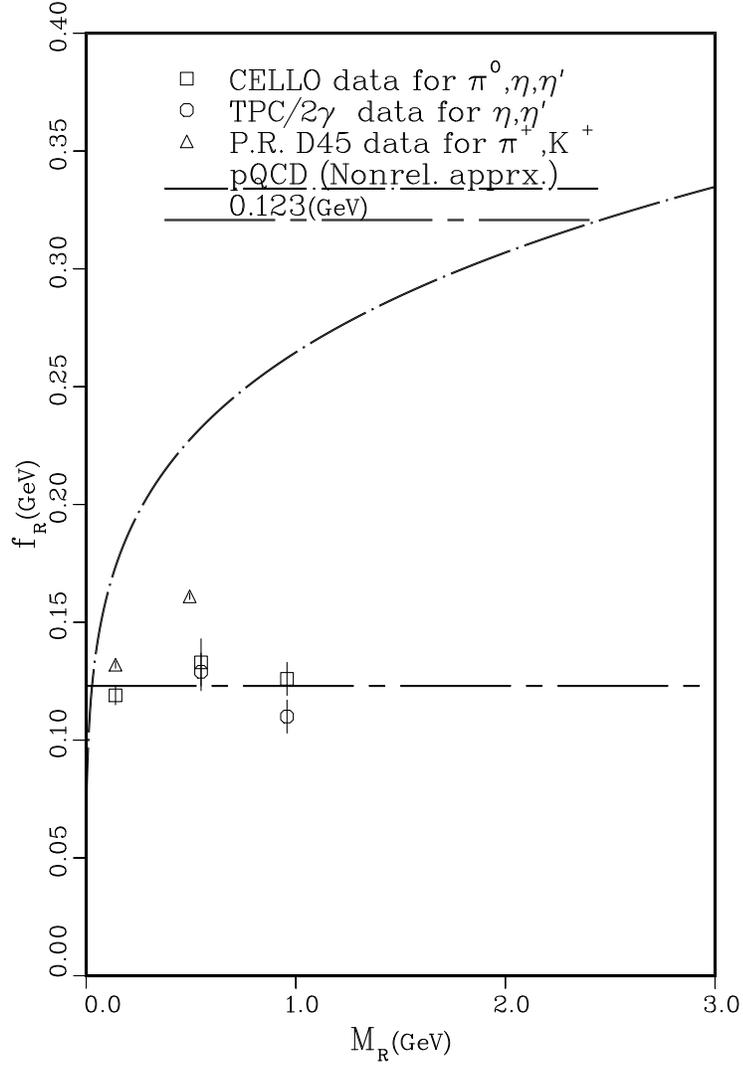}
\caption{$f_R$ for nonet pseudoscalars, and prediction from
non-relativistic potential model.}
\label{ffac.ps}
\end{figure}

\end{document}